\documentclass[aps,prb,showpacs,raggedbottom,nobalancelastpage,amssymb]{revtex4}
\usepackage{amsmath}
\usepackage{amsfonts}
\usepackage{graphicx}
\usepackage{appendix}
\usepackage{dsfont}
\usepackage{amssymb}
\usepackage{bm}
\usepackage{color}
\usepackage{ulem}
\usepackage{soul}
\usepackage{natbib}
\usepackage{nccmath}
\usepackage{array}

\newcommand{\beq}{\begin{equation}}
\newcommand{\eneq}{\end{equation}}
\newcommand{\be}{\begin{equation}}
\newcommand{\ee}{\end{equation}}
\newcommand{\bea}{\begin{eqnarray}}
\newcommand{\eea}{\end{eqnarray}}

\usepackage{multirow}
\usepackage{wasysym}


\begin{document}
\title{Transfer matrix approach to the persistent current in quantum rings: Application to hybrid normal-superconducting rings}
\author{Andrea Nava$^{1}$, Rosa Giuliano$^{1}$, Gabriele Campagnano$^{2}$, and 
Domenico Giuliano$^{1}$ }
 \affiliation{$^{1}$
Dipartimento di Fisica, Universit\`a della Calabria Arcavacata di Rende I-87036, Cosenza, Italy
and
I.N.F.N., Gruppo collegato di Cosenza, Arcavacata di Rende I-87036, Cosenza, Italy\\
$^{2}$ CNR-SPIN, Monte S. Angelo-Via Cintia, I-80126, Napoli, Italy and 
Dipartimento di Fisica, Universit\`a di Napoli ``Federico II'', 
Monte S. Angelo-Via Cintia, I-80126 Napoli, Italy}
\date{\today}

\begin{abstract}

 Using the properties of the transfer matrix of one-dimensional quantum mechanical systems, 
we derive an exact formula for the persistent current across a quantum mechanical ring pierced by a magnetic 
flux $\Phi$  as a single integral of a known function of the system's parameters. Our approach provides exact results at zero temperature,
which can be readily extended to a finite temperature  $T$. We apply our technique to  
exactly compute the  persistent current through
p-wave and s-wave superconducting-normal hybrid rings, deriving full plots of the current as 
a function of the applied flux at various system's scales. Doing so, we recover at once a number
of effects such as the crossover in the current periodicity on increasing the size of the ring 
and the  signature of the topological  phase transition in the p-wave case. 
In the limit of a large ring size,   resorting to a systematic expansion in inverse powers of the 
ring length, we derive  exact  analytic closed-form formulas, applicable to 
a number of cases of physical interest.  
\end{abstract}


\maketitle
\section{Introduction}
\label{intro}

Due to the zero-resistance of a superconductor, once a current is induced for instance by 
an applied magnetic flux $\Phi$,  in a superconducting ring, it is expected to last in principle
forever. The existence of such a persistent current  
in superconducting rings was  already predicted at the discovery of superconductivity in 1911
\cite{thinkam_1969,imry_2002}, and  later experimentally demonstrated\cite{mills}.
Remarkably, as predicted in 1983 by   B\"uttiker, Landauer and 
Imry, superconductivity is not a necessary mechanism to have persistent currents.
Indeed, even a normal ring threaded by a magnetic flux can
host a persistent current,  provided that the temperature
is low enough to suppress  inelastic scattering from phonons and other electrons
and that the size of the ring is short enough compared to the
phase coherence length \cite{bli}. Because of the gauge invariance, both for superconducting and  
normal rings, the persistent current $ I [ \Phi ]$ must be a periodic
function of the magnetic flux $\Phi$. This prediction has been eventually verified in a number of
experiments\cite{gold_experimental,aluminium_experimental}.
However,  while in the superconducting case the period is 
equal to the flux quantum appropriate for a superconductor, with current
carried by Cooper pairs, $\Phi_0^* = h   / (2e)$, 
in the normal case such a period is doubled, and equal to $\Phi_0 = h  / e$.

A large amount of literature about persistent current in a 
normal mesoscopic ring has addressed a number of issues such as
the effect of disorder in the ring with consequent possible halving 
in the period of the current \cite{meso_disorder_1,meso_disorder_2}, 
the role of the spin degree of freedom \cite{meso_spin}, the consequences 
of the electron-electron interaction, with and without impurity scattering 
\cite{meso_int}, and the presence of spin-orbit interaction \cite{meso_soi} (for a recent 
 comprehensive review  on electron  transport in mesoscopic rings see, for instance, [\onlinecite{maiti}]).
 Moreover, the issue of how the current is affected by collective fluctuations in 
quasi one-dimensional superconducting rings with a weak link has been considered \cite{giuliano_sodano_boundary},
together with the possibility of using small superconducting rings, realized with pertinent
Josephson junction networks, as   high-coherence quantum devices \cite{Cirillo,giuliano_sodano_impurity,giuliano_sodano_competing}.

Recent progresses in the fabrication of nanostructures 
made it possible to engineer hybrid devices where superconductivity is induced by
proximity effect  only in a section of the ring\cite{rings} (Hybrid Rings (HRs)). This enables one
to explore a number of  physical regimes. For instance,  one may think of looking at 
the persistent current across the ring  
varying the lengths of the two regions, while keeping the  length of the normal
region $\ell_N$  shorter than the phase coherence length. In this 
way, one may  monitor the crossover between the normal mesoscopic regime, in 
which the length of the superconducting region $\ell_S$ is shorter than
the superconducting coherence length $\xi_0$, towards the 
complementary Josephson-junction regime, in which $\ell_S \gg \xi_0$ \cite{montam}.
Accompanied to the above  crossover, one is also expected to 
see a crossover of the period of the persistent current, respect to the magnetic flux, between 
$\Phi_0$ (current carried by electrons), and $\Phi_0^*$
(current carried  by Cooper pairs) \cite{butti_1,montam}.

A special type of HR has recently attracted great interest, in view of 
the possibility of designing a non ambiguous way of probing 
Majorana fermions (MFs) in condensed matter systems. MFs have been
predicted by Kitaev to emerge as localized end modes in a 
spinless p-wave one-dimensional superconductor in its topological phase \cite{kita}.
It has been proposed that such a system can be realized
by inducing superconductivity by proximity effect in a semiconducting  quantum wire
with a  sizable spin-orbit coupling (e.g., an InAs wire) 
when subject to an external magnetic field \cite{Lutchyn,oreg}.

A zero-bias peak in a tunneling spectroscopy measurement has been
claimed as an evidence for the existence of the localized MF \cite{mourik}, 
but alternative possible explanations of the experimental data of 
Ref.[\onlinecite{mourik}] have been provided, leaving still as a debated
question whether a MF has been really detected, or not \cite{deb_1,deb_2}. 
Other proposals   to  detect MFs have been presented,
for instance, using a quantum switch made with 
two quantum dots coupled to MFs \cite{sodano_1} or by means of a local flux measurement in 
a topological rf-SQUID with a frustrating $\pi$-junction \cite{lucignano} or, ultimately, of
the analysis of the persistent current in metallic rings interrupted 
by a Coulomb blockaded topological superconducting segment \cite{buttiker}.

  When using $ I [ \Phi ]$ as a tool to monitor the emergence of MFs, 
it is of the utmost importance to disentangle effects related to MF physics from 
''spurious'' effects, which are expected to appear in nontopological phases, as well.
For instance, the crossover  of the period of $I [ \Phi ]$ {\it vs.} $\Phi$ 
from $\Phi_0$ to $\Phi_0^*$, can be either  attributed to MFs, or can be 
regarded as a simple crossover of a hybrid ring toward the mesoscopic regime, 
taking place when the length of the superconducting regions becomes 
longer than the corresponding superconducting coherence length \cite{pientka,montam}.
Therefore, it is extremely useful to recover an exact (or pertinently approximated) 
formula for $ I [ \Phi ]$, allowing us to make rigorous predictions on its dependence on
the system parameters, on the length of the superconducting and nonsuperconducting 
regions, etc.    Nevertheless, even after a number of simplifications: 
considering a ballistic system, using  a non self-consistent 
model for the superconducting region \cite{btk}, or its corresponding lattice version
for a s-wave \cite{acz}, or for a p-wave superconductor \cite{kita}, computing 
$ I [ \Phi ]$ is typically still quite a challenging task. In fact, the ''standard'' 
approach to the problem consists in computing the current as 

\beq
I [ \Phi ] = e \partial_\Phi {\cal F} [ \Phi ; T ]
\;\;\;\; , 
\label{one}
\eneq
\noindent
with ${\cal F}[ \Phi ; T ]$ being the system's free energy at applied flux $\Phi$ and 
temperature $T$. At $T= 0 $ Eq.(\ref{one}) yields 

\beq
I [ \Phi ] = e \partial_\Phi E_{\rm GS}  [ \Phi  ] =  e \partial_\Phi \sum_{E_n<0}
E_n 
\;\;\;\; , 
\label{two}
\eneq
\noindent
 where the sum is taken over energies of the occupied single-quasiparticle states. 
To compute these energies, one has to solve the secular equation for the energy 
eigenvalues at nonzero $\Phi$ with periodic boundary conditions over the whole ring. 
In general, the resulting set of equations looks quite formidable and hard to 
deal with, which requires resorting to various approximations, such as retaining only 
the low-energy part of the spectrum \cite{butti_1}, or employing various approximations
for the single-quasiparticle energies as a function of $\Phi$ in various regions of 
the spectrum \cite{montam}. 

In this paper we present a technique that, under very general assumptions such as the 
ones listed above, allows for exactly expressing $ I [ \Phi ]$ as a single integral 
of a pertinently constructed function of the system's parameters.   At $T=0$, our  approach is based on 
first writing Eqs.(\ref{two}) in terms of a single integral over an appropriate 
path in the complex energy plane, and on eventually deforming the integration
path to the imaginary axis. In this respect, our method can be regarded as an 
adapted version of the technique developed in Refs.[\onlinecite{been_0,furus,been_x2,giu_af1}] to compute
the dc Josephson current across an SNS-junction in terms of the single-quasiparticle $S$-matrix of
the junction,  which has been successfully applied to 
a number of cases of interest, such as the Josephson current through a chaotic Josephson junction \cite{brouwer},
the critical supercurrent in the quantum spin-Hall effect \cite{been.1}, and the current across a 
long Josephson junction \cite{giu_af1,giu_af2}. In fact, our method combines the 
idea of deforming the integration path  to the imaginary axis  with the idea of using the 
transfer matrix, rather than the $S$-matrix, to recover the quasiparticle scattering 
dynamics of the system. Indeed, while in principle the transfer matrix and the $S$ matrix can 
be used on an equivalent footing to derive the transport properties of a mesoscopic system 
\cite{mello_book}, the former approach comes out to be more appropriate for systems with periodic 
boundary conditions, such as quantum rings (see, for instance, Ref.[\onlinecite{chen}] for an
example of application of transfer matrix approach to the persistent current through a disordered, normal mesoscopic ring).
Besides providing an exact formula allowing to easily compute  $ I [ \Phi ]$ by numerically integrating a known 
function at fixed $\Phi$, in the large-ring limit, 
our technique is also suitable for a systematic expansion in inverse powers of the system's
length, which is the counterpart for a HR of the expansion in inverse powers of the 
length of a long SNS junction discussed in Ref.[\onlinecite{giu_af1}] for a single-channel system and 
generalized in Ref.[\onlinecite{giu_af2}] to the multi-channel case. In this limit the formula
for $ I [ \Phi ]$ is greatly simplified, allowing for the derivation of 
analytic closed-form formulas for the current in a number of cases of interest.
As an example of the effectiveness of our technique, we consistently extend the results 
of Refs.[\onlinecite{buttiker,pientka,choi}] by  deriving exact   plots of 
$ I [ \Phi ]$ {\it vs.} $\Phi$ through
p-wave and s-wave superconducting-normal hybrid rings, which, by comparing the plots in the two 
cases with each other, allows us to  highlight the features strictly related to the emergence of 
MFs in the p-wave case.

Remarkably, while in this paper we do not  account for a number of features that can be
in principle important in HRs, such as quantum phase slips \cite{slips}, 
electronic interaction effects \cite{meso_int}, disorder \cite{meso_disorder_1,meso_disorder_2}, 
etc., as we outline among the 
concluding remarks, it should not be difficult to pertinently address them within our formalism.
In fact,  we plan to treat some
of this issues in forthcoming publications, as a natural development of this work.  

The paper is organized as follows:

\begin{itemize}
 \item In section \ref{transfer_persist} we introduce the transfer matrix approach which we 
 use  to compute the persistent current;
 
 \item In section \ref{modeham} we introduce the lattice model Hamiltonian for a hybrid superconducting-superconducting
 ring, both for the p-wave and for the s-wave superconductor;
 
 \item In section \ref{calc_persist} we compute the persistent current in some relevant reference model, such as 
 a superconducting ring interrupted by a normal weak link, and in the case of a hybrid normal-superconducting 
 ring. Specifically, we recover both cases as particular limit of the model introduced in section \ref{modeham}:
 the former one by making one of the two superconducting region shrink to zero, the latter one by simply setting to 
 zero the superconducting gap in one of the two regions;
 
 \item In section \ref{large} we consider the limit of large ring size in the HR both in the s-wave, and 
 in the p-wave case. In particular, we first show how one 
 recovers the results of Refs.[\onlinecite{giu_af1,giu_af2}] in the limit of infinite length for the superconducting region, 
 therefore, we discuss the complementary limit of a short-superconducting and a long-normal region;
 
 \item In section \ref{finite_T} we outline how our results can be extended to a temperature $T$ finite, but well below 
 the superconducting gap. Specifically, this allows us to recover within our formalism the finite-$T$
 formula at Eq.(\ref{one});
 
 \item In section \ref{conclusions} we provide our main conclusions and discuss further possible developments of our work;
 
 \item In appendix \ref{exw} we review the derivation of the  sub-gap states 
 in an open Kitaev chain, of finite length $\ell_S$, which is functional to the discussion of 
 the results of section \ref{calc_persist}.
 
\end{itemize}

\section{The transfer matrix approach to the persistent current}
\label{transfer_persist}
 
Our technique to compute the persistent current through mesoscopic normal/superconducting rings 
is based on a combination of the approach to  the Josephson current across an SNS junction based 
on the analytical continuation of the quasiparticle $S$-matrix to the imaginary axis \cite{been_0,furus,been_x2,giu_af1} 
with the transfer matrix (TM) approach to transport in mesoscopic systems \cite{mello_book}, particularly well-suited to 
account for periodic boundary conditions in quantum rings. The key feature of our 
approach is that it eventually leads to an exact, closed-form formula for the groundstate energy of the system at 
a given $\Phi$ and, therefore, for the persistent current 
across the ring.  In order to illustrate its main features,  in this section we review the main steps 
leading to our final formula for the groundstate energy in the case of a normal, mesoscopic
ring. Nevertheless, as we discuss in the following, the case of a HR 
containing one, or more, superconducting sections  is quite a straightforward generalization of 
the derivation we provide in this section. To treat the p-wave and the s-wave case on 
the same footing we choose to perform our derivation within a lattice one-dimensional
model Hamiltonian, which, in the superconducting region, 
corresponds to Kitaev's Hamiltonian in the p-wave case \cite{kita}, and to the 
lattice one-dimensional Hamiltonian of Ref.[\onlinecite{acz}] in the s-wave case.

\begin{figure}
\includegraphics*[scale=0.5]{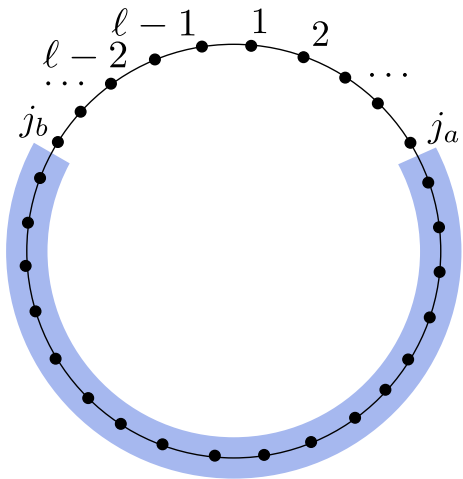}
\caption{An example of the ring geometry used in the article: nontrivial scattering
happens in the shaded region whereas in the remaining part one can write the wave
function in terms of simple scattering states (cfr. main text)}
\label{ring-example}
\end{figure}
As an introduction to the technique presented in this article, let us consider a normal (non superconducting) ring
with $\ell$ sites as depicted in Fig.\ref{ring-example}. We can ideally divide the ring into three regions, within the shaded region 
non-trivial scattering processes may happen, while in the remaining part the system is described by the  
asymptotic lattice Hamiltonian of the form $H_{\rm as} = - J \sum_j \{ c_j^\dagger 
c_{ j + 1 } + c_{ j +1}^\dagger c_j \} - \mu \sum_j c_j^\dagger c_j $, with $c_j$ being the single-fermion 
lattice annihilation operator for spin-less fermions. 

Let us for the moment imagine that the ring is open at the last site $\ell$ and that $j_1$ and $j_2$ are limiting the 
 region of the ring with non trivial scattering.  The wave function at energy $E$ in the two "asymptotic"  
regions for $j < j_1$ and for $j>j_2$ is 

\begin{eqnarray}
 u_j &\sim& A_+^< e^{ i k j } + A_-^< e^{ - i k j } \:\: , \: (j < j_1 )  \nonumber \\
  u_j &\sim& A_+^> e^{ i k j } + A_-^> e^{ - i k j } \:\: , \: (j > j_2 )
  \:\:\:\: , 
  \label{ra.1}
\end{eqnarray}
\noindent
with 

\beq
E = - 2 J \cos ( k ) - \mu
\:\:\:\: . 
\label{ra.2}
\eneq
\noindent
Now, by definition the transfer matrix between sites  $j_a < j_1$  and  $j_b > j_2$, ${\cal M} [ E ; j_a , j_b ]$,  relates 
the solution at $j = j_a$ to the solution at $j = j_b$, that is 

\beq
u_{ j_b } = \tilde{A}_+ e^{ i k j_a } + \tilde{A}_- e^{ - i k j_a } 
\;\;\;\; , 
\label{ra.3}
\eneq
\noindent
with 

\beq
\left[ \begin{array}{c}
        \tilde{A}_+ \\ \tilde{A}_- 
       \end{array} \right] = {\cal M} [ E ; j_a , j_b ]
\left[ \begin{array}{c}
        A_+^< \\ A_-^<
       \end{array} \right] 
\:\:\:\: . 
\label{ra.4}
\eneq
\noindent
Upon considering the closed geometry, we have to impose periodic boundary conditions (PBCs) on 
the solution in Eq.(\ref{ra.1}). Going through Eq.(\ref{ra.3}), this is accounted for 
by means of the  secular equation 

\beq
{\cal M} [ E ; 1 , \ell  ] \left[ \begin{array}{c}
        A_+^< \\ A_-^<
       \end{array} \right]  = \left[ \begin{array}{c}
        A_+^< \\ A_-^<
       \end{array} \right]
\:\:\:\: . 
\label{ra.5}
\eneq
\noindent
As a result, Eq.(\ref{ra.5}) leads to the secular equation for the allowed values of 
the energy $E$:

\beq
 {\rm det} \{ {\cal M}  [ E ; \ell  ] - {\bf I}  \} = 0 
\:\:\:\: . 
\label{ra.6}
\eneq
\noindent
 (In Eq.(\ref{ra.6}) we suppressed  the dependence of  ${\cal M} [ E ; j_a , j_b  ]$ on $j_a$ since, 
as expected because of the periodic boundary conditions, it depends only on 
the distance between the sites, which is equal  to $ \ell$.)   
Besides constituting an alternative way of presenting the eigenvalue equation for 
a single particle on the ring, Eq.(\ref{ra.6}) also provides an efficient way to
compute the   groundstate energy of the system, $E_{\rm GS}$, defined as the sum of all 
the negative (if measured with respect to the chemical potential) 
single-particle energy eigenvalues $E_n$, that is

\beq
E_{\rm GS} = \sum_{ E_n < 0} E_n 
 \:\:\:\: . 
 \label{ra.7}
 \eneq
 \noindent
 In fact, single-particle negative energy eigenvalues 
 are just the zeros of $ {\rm det} \{ {\cal M}  [ E ; \ell  ] - {\bf I} \} $ lying at the 
 negative part of the real axis, if one regards $ {\rm det} \{ {\cal M}  [ E ; \ell  ]- {\bf I}  \} $ as  a function of the 
 complex variable $E$. To sum over all of them, we adapt the approach of Refs.[\onlinecite{giu_af1,giu_af2}], namely, 
 we first of all note that  the energy eigenvalues are the poles (with residues all equal to 1)
 of the function $\Psi [ E ; \ell ]$, defined as 

\beq
\Psi [ E ; \ell ] = \frac{ \partial_E  {\rm det} \{ {\cal M}  [ E ; \ell  ] - {\bf I} \} }{ 
 {\rm det} \{ {\cal M}  [ E ; \ell  ] - {\bf I}  \} } = 
 \partial_E \ln  {\rm det} \{ {\cal M}  [ E ; \ell  ] - {\bf I} \}
 \:\:\:\: . 
 \label{ra.8}
 \eneq
 \noindent
 
\begin{figure}
\includegraphics*[width=1\linewidth]{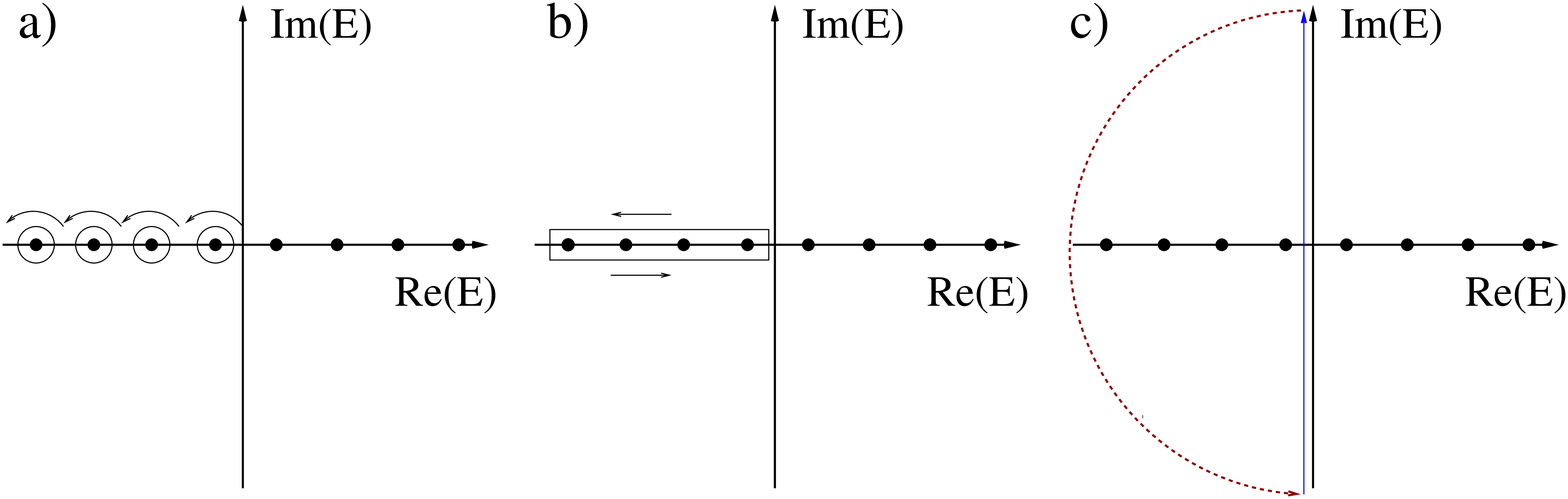}
\caption{Sequence of deformations in the integration path $\Gamma$ eventually 
allowing to express $ I [ \Phi ]$ as an integral over the imaginary axis: \\
{\bf a)} The path $\Gamma$ obtained as the union of small circles, each one 
surrounding one, and only one, negative (with respect to the Fermi level) energy
eigenvalue; \\
{\bf b)} The integral over $\Gamma$ can be deformed to an integral over just one closed path, 
surrounding all the negative energy eigenvalues; \\
{\bf c)} The integral over the red-dashed arc is assumed to be equal to 0 in the infinite-radius
limit. Only the integral over the imaginary axis (to which the solid-blue line can be 
continuously deformed) is left.} \label{intpath}
\end{figure}
\noindent
We therefore introduce the integration path $\Gamma$ depicted in Fig.\ref{intpath}{\bf a)} 
and compute $E_{\rm GS}$ as 

\beq
E_{\rm GS} = \frac{1}{2 \pi i } \: \oint_\Gamma \: d E \: E \Psi [ E ; \ell ] 
= -  \frac{1}{2 \pi i } \: \oint_\Gamma \: d E \:  \ln  {\rm det} \{ {\cal M}  [ E ; \ell  ]  - {\bf I} \}
\:\:\:\: . 
\label{ra.9}
\eneq
\noindent 
On noting that  $\Psi [ E ; \ell ] $ must have no singularities outside of the real
axis, we deform the integration path as illustrated in Fig.\ref{intpath}, so to eventually
compute the integral over the imaginary axis as 

\beq
E_{\rm GS} =  -  \frac{1}{2 \pi  } \:  \int_{ - \infty}^{  \infty} 
\: d \omega \:  \ln  {\rm det} \{ {\cal M}  [ i \omega  ; \ell  ]  - {\bf I} \}
\:\:\:\: . 
\label{ra.10}
\eneq
\noindent
Equation (\ref{ra.10}) is the heart of our article. When the ring is pierced by a magnetic field 
the transfer matrix $ {\cal M}$ and hence the spectrum    takes an additional dependence  on the total 
flux $\Phi$. We can calculate   the (zero-temperature) persistent current as 
 
\beq
 I [ \Phi ]  = e \partial_\Phi E_{\rm GS} [ \Phi ] =  -  \frac{e}{2 \pi  } \:  \int_{ - \infty} ^{  \infty} 
\: d \omega \:  \partial_\Phi \ln  {\rm det} \{ {\cal M}  [ i \omega  ; \ell  ; \Phi  ]- {\bf I}  \}
\:\:\:\: . 
\label{ra.11}
\eneq
\noindent  
Equation (\ref{ra.11}), together with the  corresponding finite-$T$ generalization which we 
discuss in the following sections, is the key result of our paper and its main point of novelty: 
it provides an exact formula for $I [ \Phi ]$ which can be readily implemented, once one
has derived the TM of the ring. Remarkably, while, for the sake of the presentation, in 
this section we relied on a normal mesoscopic ring, Eq.(\ref{ra.11}) does certainly apply equally well
to the case of superconducting and/or hybrid rings, once one has constructed the appropriate 
TM for the system. Importantly enough, our approach allows for circumventing 
  the main difficulty in  computing the persistent current working out the single-particle energy levels, using 
them to compute $E_{\rm GS} [ \Phi ]$, and eventually employing the final result to derive the current. Besides 
 a few, oversimplified remarkable exceptions, extracting the single-quasiparticle 
energy levels from the TM (or by means of  an equivalent approach) is quite a difficult task to achieve, 
which can only be addressed either by means of exact numerical diagonalization of the Hamiltonian (with the 
corresponding limitations on the system size), or within approximate methods based on truncating 
the full Hilbert space to a small number of low-energy (sub-gap) states, which are expected to carry the largest part 
of the current. While both these techniques are expected to suffer of strong limitations in their range of applicability,  
our Eq.(\ref{ra.11}) is universally applicable and, once one has constructed the TM, does not require to go 
through any additional step, besides explicitly computing the integral at the right-hand side. 

Our Eq.(\ref{ra.11}) is the analog, for a superconducting ring, of the equation giving the Josephson current 
across an SNS junction in terms of the single-particle $S$-matrix, analytically continued to the imaginary 
axis \cite{been_0,been_x2}. In this respect, it is expected by analogy to have a wide range of applicability, either 
as it stands, as an exact, closed form formula for $I [ \Phi ]$, or as a starting point to account, for instance, 
for the effect of disorder (in analogy to what is done in Ref.[\onlinecite{brouwer}] for an SNS junction), or 
of the electronic interaction. As an example of application of our technique, in the following we
derive full  plots of the persistent current in hybrid superconducting-normal rings, both in 
the case of p-wave, and of s-wave superconductivity, which allows us to highlight the key feature that, in 
the former case, can be definitely attributed to the emergence of MFs at the interfaces. 

As a convention on the units of measurement, in the following 
  we shall  measure the flux $\Phi$ in units of $\hbar   / (2 e) $,
so that a period $\Phi_0^*$ corresponds to a $2 \pi$-periodicity, while a 
period $\Phi_0$ corresponds to a $4 \pi$-periodicity.

\section{Model Hamiltonian for a hybrid superconducting-normal ring}
\label{modeham}

To model a hybrid superconducting-normal ring, we resort to an effective lattice model
Hamiltonian with position-dependent parameters. Using a lattice model Hamiltonian allows, on 
one hand, to easily introduce p-wave pairing both in real space and in momentum space, on the other
hand, it provides a natural mean to regularize divergences which would arise in the continuum 
model when summing over the energies of the occupied levels
to compute $E_{\rm GS} [ \Phi ]$ (and, therefore, $I [ \Phi ]$). To 
highlight special features of p-wave HRs, in the following, when possible, 
we systematically compare the results obtained in p-wave systems to the ones 
obtained in s-wave systems. Therefore,   
in this section we  derive  the transfer matrix which we shall eventually use in the following to 
compute the persistent current in both cases. 

\subsection{Model Hamiltonian for  a p-wave nonhomogeneous superconducting ring}
\label{mpwave}

\begin{figure}
\includegraphics*[width=.6\linewidth]{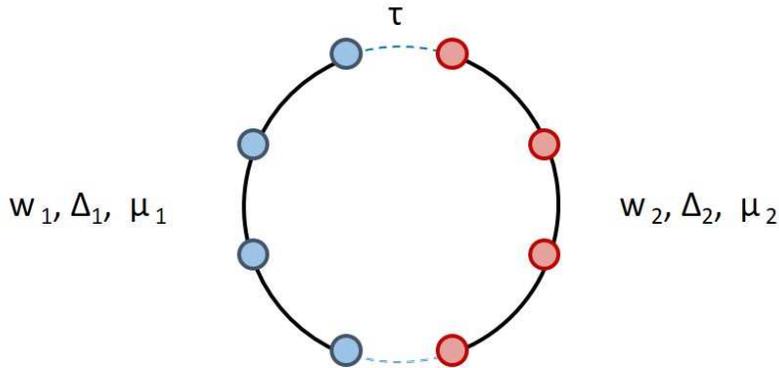}
\caption{Sketch of the system described by the model Hamiltonian in Eq.(\ref{pw.1}).} \label{hyring}
\end{figure}

We here consider  the system sketched in Fig.\ref{hyring}: a
hybrid ring made of two homogeneous regions, either superconducting (p-wave or s-wave), or 
normal, each one characterized by different parameters. In the p-wave case,  referring to 
the one-dimensional Kitaev Hamiltonian \cite{kita} for a spinless p-wave superconductor, 
calling 1 and 2 the two regions, we assume that the normal hopping amplitude, the 
pairing gap and the chemical potential are respectively given by $w_1 , \Delta_1 , \mu_1$ and 
by $w_2 , \Delta_2 , \mu_2$. Moreover, we assume that regions 1 and 2 are coupled at their endpoints via a normal
hopping term, with hopping amplitude $\tau$. In a ring configuration, we also assume that a magnetic
flux $\Phi$ pierces the ring. By means of  an appropriate canonical redefinition of the lattice fermion operators, 
it is possible to account for  the applied magnetic flux 
in terms of a phase factor $e^{ \pm i \frac{\Phi}{4}}$, symmetrically ascribed to  
the two hopping terms between the two 
regions. As a result, we eventually present  the model Hamiltonian as

\begin{eqnarray}
 H &=& - w_1 \sum_{ j = 1}^{ \ell_1 - 1 } \{ c_j^\dagger c_{ j + 1 } + 
 c_{ j + 1}^\dagger c_j \} - \mu_1 \sum_{ j = 1}^{\ell_1 } c_j^\dagger c_j 
 + \Delta_1 \sum_{ j = 1}^{ \ell_1 - 1 } \{ c_j c_{ j + 1 } + c_{ j + 1}^\dagger 
 c_j^\dagger \}
 \nonumber \\
 &  - & w_2 \sum_{ j = 1}^{ \ell_2 - 1 } \{ d_j^\dagger d_{ j + 1 } + 
 d_{ j + 1}^\dagger d_j \} - \mu_2 \sum_{ j = 1}^{\ell_2 } d_j^\dagger d_j 
 + \Delta_2 \sum_{ j = 1}^{ \ell_2 - 1 } \{ d_j d_{ j + 1 } + d_{ j + 1}^\dagger 
 d_j^\dagger \}
 \nonumber \\
 &  - &\tau \{ [ c_1^\dagger d_{ \ell_2} + d_1^\dagger c_{ \ell_1 } ] e^{ \frac{i}{4} \Phi } 
 + [ d_{ \ell_2}^\dagger c_1 + c_{ \ell_1 }^\dagger d_1 ] e^{  - \frac{i}{4} \Phi }  \}
 \:\:\:\: . 
 \label{pw.1}
\end{eqnarray}
\noindent
\noindent
In Eq.(\ref{pw.1}) we have set the lengths of the two arms of the ring respectively 
at $\ell_1$ and at $\ell_2$ and used $c_j , c_j^\dagger$ to denote the lattice annihilation/creation operators at 
site $j$ within region 1, and $d_j , d_j^\dagger$ to denote the lattice annihilation/creation operators at 
site $j$ within region 2, with standard anticommutation relations $\{ d_j , d_{j'}^\dagger \} 
= \{ c_j , c_{j'}^\dagger \} = \delta_{ j , j'}$, all the other anticommutators being 
equal to 0. Taken in the appropriate limit, the model Hamiltonian in Eq.(\ref{pw.1}) is suitable to 
describing a number of systems of physical interest, such as a p-wave superconducting ring interrupted
by a weak link \cite{choi}, a hybrid p-wave-normal metal ring, and a half-topological, half-non-topological
superconducting ring \cite{pientka}. Moreover, as we discuss in the following, taken in the limit of
long-superconducting section, the superconducting-normal hybrid ring is mapped onto the 
effective model for a Josephson junction made with topological superconductors \cite{pikulin, giuaf_3}
 
In the following, we will be mostly focusing onto the $\Delta_2 \to 0$ limit. 
To construct the transfer matrix, we start from   the  
Bogoliubov-de Gennes (BdG) equations for the single-quasiparticle wavefunction at a given energy $E$. 
To do so, we consider a generic energy eigenmode $\Gamma_E$ which, in terms of the single-fermion lattice 
operators on the ring, is given by 

\beq
\Gamma_E = \sum_{ j = 1}^{ \ell_1 } \{ [ u_j^{(1)} ]^* c_j +  [ v_j^{(1)} ]^* c_j^\dagger \}
+  \sum_{ j = 1}^{ \ell_2 } \{ [ u_j^{(2)} ]^* d_j +  [ v_j^{(2)} ]^* d_j^\dagger \}
\:\:\:\: , 
\label{pw.2}
\eneq
\noindent 
with $\left[ \begin{array}{c}
              u_j^{(1)} \\ v_j^{(1)} 
             \end{array} \right]$ and  $\left[ \begin{array}{c}
              u_j^{(2)} \\ v_j^{(2)} 
             \end{array} \right]$ being the single-quasiparticle 
             wavefunction in region-1 and in region-2.
On imposing the canonical commutation relation 

\beq
[ \Gamma_E , H ] = E \Gamma_E 
\:\:\:\: , 
\label{pw.3}
\eneq
\noindent
we therefore obtain  the BdG  equations for the wavefunctions. Within the 
homogeneous regions, these are given by 

\begin{eqnarray}
 E u_j^{(1)} &=& - w_1 \{ u_{j+1}^{(1)} + u_{ j - 1}^{(1)} \} 
 - \mu_1 u_j^{(1)} + \Delta_1 \{ v_{j+1}^{(1)} - v_{ j - 1}^{(1)} \} \nonumber \\
  E v_j^{(1)} &=&  w_1 \{ v_{j+1}^{(1)} + v_{ j - 1}^{(1)} \} 
 + \mu_1 v_j^{(1)} - \Delta_1 \{ u_{j+1}^{(1)} - u_{ j - 1}^{(1)} \}
 \:\:\:\: , 
 \label{pw.4}
\end{eqnarray}
\noindent
for $1 < j < \ell_1$, and

\begin{eqnarray}
 E u_j^{(2)} &=& - w_2 \{ u_{j+1}^{(2)} + u_{ j - 1}^{(2)} \} 
 - \mu_2 u_j^{(2)} + \Delta_2 \{ v_{j+1}^{(2)} - v_{ j - 1}^{(2)} \} \nonumber \\
  E v_j^{(2)} &=&  w_2 \{ v_{j+1}^{(2)} + v_{ j - 1}^{(2)} \} 
 + \mu_2 v_j^{(2)} - \Delta_2 \{ u_{j+1}^{(2)} - u_{ j - 1}^{(2)} \}
 \:\:\:\: , 
 \label{pw.5}
\end{eqnarray}
\noindent
for $1 < j < \ell_2$. At the interfaces between the two regions, instead, 
the BdG equations yield

\begin{eqnarray}
 E u_1^{(1)} &=& - w_1 u_2^{(1)} - \tau e^{ - \frac{i}{4 } \Phi } u_{ \ell_2}^{(2)} 
 - \mu_1 u_1^{(1)} + \Delta_1 v_2^{(1)} \nonumber \\
 E v_1^{(1)} &=&   w_1 v_2^{(1)} + \tau e^{  \frac{i}{4 } \Phi } v_{ \ell_2}^{(2)} 
 + \mu_1 v_1^{(1)} - \Delta_1 u_2^{(1)} \nonumber \\
 E u_{\ell_2}^{(2)} &=& - w_2 u^{(2)}_{\ell_2 - 1 } - \tau e^{ \frac{i}{4} \Phi } u_1^{(1)} 
 - \mu_2 u_{ \ell_2}^{(2)} - \Delta_2 v_{ \ell_2 - 1 }^{(2)} \nonumber \\
  E v_{\ell_2}^{(2)} &=&  w_2 v^{(2)}_{\ell_2 - 1 } + \tau e^{ - \frac{i}{4} \Phi } v_1^{(1)} 
 + \mu_2 v_{ \ell_2}^{(2)} + \Delta_2 u_{ \ell_2 - 1 }^{(2)} \nonumber \\
 E u_1^{(2)} &=& - w_2 u_2^{(2)} - \tau e^{ - \frac{i}{4} \Phi } u_{ \ell_1}^{(1)} - \mu_1 
 u_1^{(2)} + \Delta_2 v_2^{(2)} \nonumber \\
  E v_1^{(2)} &=&  w_2 v_2^{(2)} + \tau e^{  \frac{i}{4} \Phi } v_{ \ell_1}^{(1)} + \mu_1 
 v_1^{(2)} - \Delta_2 u_2^{(2)} \nonumber \\
 E u_{ \ell_1}^{(1)} &=& - w_1 u_{ \ell_1 - 1 }^{ (1)} - \tau e^{ \frac{i}{4} \Phi } u_1^{(2)} 
 - \mu_1 u_{ \ell_1}^{(1)} - \Delta_1 v_{ \ell_1 - 1}^{(1)} \nonumber \\
  E v_{ \ell_1}^{(1)} &=&   w_1 v_{ \ell_1 - 1 }^{ (1)} + \tau e^{-  \frac{i}{4} \Phi } v_1^{(2)} 
 + \mu_1 v_{ \ell_1}^{(1)} + \Delta_1 u_{ \ell_1 - 1}^{(1)}
 \:\:\:\: . 
 \label{pw.6}
\end{eqnarray}
\noindent
According to Eqs.(\ref{pw.4},\ref{pw.5}), within the homogeneous regions, 
we write the wavefunctions as superpositions of the solutions of the 
homogeneous BdG equations, that is, we set  

\beq
\left[ \begin{array}{c}
        u_j \\ v_j 
       \end{array} \right] = \left[ \begin{array}{c}
                                     u^{(1)} \\ v^{(1)} 
                                    \end{array} \right] e^{ i k_1 j } 
                                    \:\:\:\: , 
                                    \label{pw.7}
\eneq
\noindent
in region 1, with  $1 \leq j \leq \ell_1$, and  
                                  
\beq
\left[ \begin{array}{c}
        u_j \\ v_j 
       \end{array} \right] = \left[ \begin{array}{c}
                                     u^{(2)} \\ v^{(2)} 
                                    \end{array} \right] e^{ i k_2 j } 
                                    \:\:\:\: , 
                                    \label{pw.8}
\eneq
\noindent
in region 2, with  $1 \leq j \leq \ell_2$. At a given energy $E$, the 
amplitudes $\left[ \begin{array}{c}
                                     u^{(a)} \\ v^{(a)} 
                                    \end{array} \right]$ ($a = 1 , 2 $)  are determined by 
solving the equations

\begin{eqnarray}
 0 &=& \{E + 2 w_a \cos ( k_a ) + \mu_a \} u^{(a)} - 2 i \Delta_a \sin ( k_a ) v^{(a)} \nonumber \\
 0 &=& 2 i \Delta_a \sin ( k_a ) u^{(a)} + \{E - 2 w_a \cos ( k_a ) -\mu_a \} v^{(a)} 
 \:\:\:\: , 
 \label{pw.9}
\end{eqnarray}
\noindent
with $k_1 , k_2$ determined by the dispersion relations

\beq
E^2 = [ 2 w_1 \cos ( k_1 ) + \mu_1 ]^2 + 4 \Delta_1^2 \sin^2 ( k_1 ) = 
 [ 2 w_2 \cos ( k_2 ) + \mu_2 ]^2 + 4 \Delta_2^2 \sin^2 ( k_2 ) 
 \:\:\:\: . 
 \label{pw.11}
 \eneq
 \noindent
Solving Eq.(\ref{pw.11}) at a given energy, we 
 define the particle-like momenta $k_p^{(a)}$ and the hole-like 
momenta $k_h^{(a)}$, as 

\begin{eqnarray}
 \cos ( k_p^{(a)}  ) &=& - \frac{w_a \mu_a}{2 ( w_a^2 - \Delta_a^2 ) }
 - \frac{1}{2} \sqrt{\frac{E^2 - [ \Delta_w^{(a)} ]^2 }{w_a^2 - \Delta_a^2}} \nonumber \\
  \cos ( k_h^{(a)}  ) &=& - \frac{w_a \mu_a}{2 ( w_a^2 - \Delta_a^2 ) }
 + \frac{1}{2} \sqrt{\frac{E^2 - [ \Delta_w^{(a)} ]^2 }{w_a^2 - \Delta_a^2}} 
\:\:\:\: , 
\label{pw.12}
 \end{eqnarray}
\noindent
with 

\beq
\Delta_w^{(a)} = \Delta_a \sqrt{4 - \frac{\mu_a^2}{w_a^2 - \Delta_a^2}}
\:\:\:\: . 
\label{pw.13}
\eneq
\noindent
(Note that Eqs.(\ref{pw.12}) do in principle hold also for $ | E | < \Delta_w^{(a)}$). 
At  given $k_p^{(a)} , k_h^{(a)}$, one determines the wavefunctions 
$ ( u_p^{(a)}  , v_p^{(a)} )$ and $ ( u_h^{(a)}  , v_h^{(a)} )$, defined as 
solutions of  Eqs.(\ref{pw.9}) with $k_a$ respectively equal to $k_p$ and $k_h$.   
Taking the most general linear combinations of 
wavefunctions at the same energy, one finds that  a generic wavefunction at energy $E$ within region-$a$ (=1,2) 
takes the form 

\beq
\left[ \begin{array}{c}
        u_j^{(a)}  \\ v_j^{(a)}  
       \end{array} \right] = 
       A_{ (p , + )}^{(a)} \left[ \begin{array}{c}
                                   u_p^{(a)} \\ v_p^{(a)} 
                                  \end{array} \right] e^{ i k_p^{(a)} j } 
                                  + 
       A_{ (p , - )}^{(a)} \left[ \begin{array}{c}
                                   u_p^{(a)} \\ - v_p^{(a)} 
                                  \end{array} \right] e^{ - i k_p^{(a)} j }                                   
 + A_{ (h , + )}^{(a)} \left[ \begin{array}{c}
                                   u_h^{(a)} \\ - v_h^{(a)} 
                                  \end{array} \right] e^{ - i k_h^{(a)} j } 
                                  + 
       A_{ (h , - )}^{(a)} \left[ \begin{array}{c}
                                   u_h^{(a)} \\ v_h^{(a)} 
                                  \end{array} \right] e^{  i k_h^{(a)} j }  
\:\:\:\: . 
\label{pw.14}
\eneq
\noindent
The transfer matrix is fully determined once one recovers the relations between the amplitudes 
$A_{ (p , + )}^{(a)} ,  A_{ (p , - )}^{(a)} ,  A_{ (h , + )}^{(a)}  , A_{ (h , - )}^{(a)} $ in 
the two regions. These are determined by the interface conditions obtained from 
Eqs.(\ref{pw.6}). In a compact notation, these are given by 
 
\beq
  \Sigma_a [ \Phi] \cdot [ \Upsilon^{(a)}  ( E ) ] \cdot \left[ \begin{array}{c}
A_{(p,+)}^{(a)} \\      A_{(p,-)}^{(a)} \\   A_{(h,+)}^{(a)}    \\   A_{(h,-)}^{(a)}                                                                
                                                              \end{array} \right] 
 =  \Omega_b [ \Phi ]  \cdot [ \Upsilon^{(b)}  ( E ) ] \cdot {\bf T}_b
 ( E  ; \ell_b ) \cdot   \left[ \begin{array}{c}
A_{(p,+)}^{(b)} \\      A_{(p,-)}^{(b)} \\   A_{(h,+)}^{(b)}    \\   A_{(h,-)}^{(b)}                                                                
                                                              \end{array} \right] 
                                                              \:\:\:\: ,
                                                              \label{pw.15}
\eneq
\noindent
with $b = 2 (1)$ if $a=1 (2)$, and the matrix $\Sigma_a [ \Phi ] $ defined as 

\beq
\Sigma_a [ \Phi ] = \left[ \begin{array}{cccc}
\tau e^{ \frac{i}{4} \Phi } & 0 & 0 & 0 \\
0 & \tau e^{ - \frac{i}{4} \Phi } & 0 & 0 \\
0 & 0 & w_a & \Delta_a \\ 0 & 0 & \Delta_a & w_a 
                           \end{array} \right] 
\;\;\;\; , 
\label{pw.16}
\eneq
\noindent
  the matrix $[ \Upsilon^{(a)}  ( E ) ] $ defined as 

\beq
[ \Upsilon^{(a)}  ( E ) ] = \left[ \begin{array}{cccc}
e^{ i k_p^{(a)} } u_p^{(a)} &    e^{ - i k_p^{(a)} } u_p^{(a)}     &
e^{ -  i k_h^{(a)} } u_h^{(a)} &    e^{  i k_h^{(a)} } u_h^{(a)} \\
e^{ i k_p^{(a)} } v_p^{(a)} &   -  e^{ - i k_p^{(a)} } v_p^{(a)}     &
 - e^{ -  i k_h^{(a)} } v_h^{(a)} &   e^{  i k_h^{(a)} } v_h^{(a)} \\
  u_p^{(a)} &      u_p^{(a)}     &
 u_h^{(a)} &     u_h^{(a)} \\
  v_p^{(a)} &   -  v_p^{(a)}     &
  - v_h^{(a)} &     v_h^{(a)} 
                                   \end{array} \right]
\:\:\:\: ,
\label{pw.17}
\eneq
\noindent
the matrix $\Omega_a [ \Phi ]$ given by 

\beq
\Omega_a [ \Phi ] = \left[ \begin{array}{cccc}
w_a & - \Delta_a & 0 & 0 \\
- \Delta_a & w_a & 0 & 0 \\
0 & 0 &  \tau e^{ - \frac{i}{4} \Phi } & 0 \\
0 & 0 & 0 &  \tau e^{  \frac{i}{4} \Phi }
                         \end{array} \right]
\:\:\:\: , 
\label{pw.18}
\eneq
\noindent
and, finally, the transfer matrix for a homogeneous region of 
length $\ell$ being given by

\beq
{\bf T}_a ( E ; \ell ) = \left[ \begin{array}{cccc}
                                 e^{ i k_p^{(a)} \ell } & 0 & 0 & 0 \\
0 &    e^{ -  i k_p^{(a)} \ell } & 0 & 0  \\
0 & 0 &  e^{ -  i k_h^{(a)} \ell } & 0   \\
0 & 0 & 0 & e^{   i k_h^{(a)} \ell }
                                \end{array} \right]
\:\:\:\: . 
\label{pw.19}
\eneq
\noindent
 As a result,  the transfer matrix for the full ring takes the form 

\beq  
{\cal M}_{ p - wave } [ E ; \Phi ; \ell_1 ; \ell_2 ]  =  [ \Upsilon^{(2)}  ( E ) ]^{-1} \cdot  \Sigma_2^{-1}  [ \Phi] \cdot \Omega_1 [ \Phi ]  \cdot [ \Upsilon^{(1)}  ( E ) ] 
\cdot {\bf T}_1 ( E  ; \ell_1)  
  \cdot [ \Upsilon^{(1)}  ( E ) ]^{-1} \cdot \Sigma_1^{-1} [ \Phi] \cdot 
 \Omega_2 [ \Phi ]  \cdot [ \Upsilon^{(2)}  ( E ) ] \cdot {\bf T}_2 ( E  ; \ell_2)  
 \:\:\:\: . 
 \label{pw.21}
 \eneq
 \noindent
Equation (\ref{pw.21}) is the key ingredient we need to compute the persistent current in the 
various regimes of interest. As stated above, we now derive the analog of  Eq.(\ref{pw.21})
 in the case of a ring made of s-wave superconducting regions.

\subsection{Model Hamiltonian for an s-wave nonhomogeneous superconducting ring}
\label{mswave}

\begin{figure}
\includegraphics*[width=.7\linewidth]{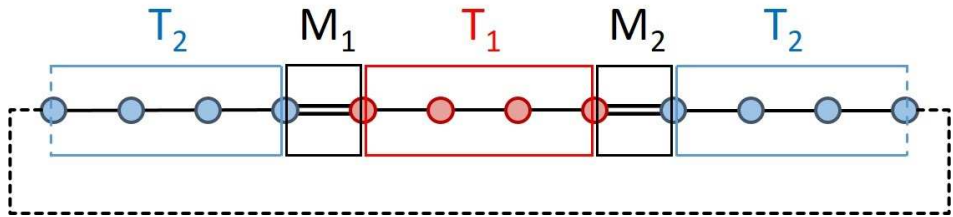}
\caption{Graphical representation of the factorizability of the transfer matrix: for the specific
system depicted in figure, the transfer matrix is given by ${\cal M} = {\bf T}_2 \cdot M_1 \cdot {\bf T}_1 
\cdot M_2 \cdot {\bf T}_2$ (from right to left).
} \label{tm}
\end{figure}

In the case of a  
system made of two s-wave superconducting regions with parameters respectively 
given by $w_1 , \Delta_1 , \mu_1$ and by $w_2 , \Delta_2 , \mu_2$, and connected to
each other with hopping amplitude $\tau$, the corresponding (spinful) Hamiltonian is 
given by

\begin{eqnarray}
 H &=& - w_1 \sum_\sigma \sum_{ j = 1}^{ \ell_1 - 1 } \{ c_{ j , \sigma}^\dagger 
 c_{ j + 1  , \sigma } +   c_{ j + 1 , \sigma }^\dagger c_{j , \sigma} \} - 
 \mu_1 \sum_\sigma \sum_{ j = 1}^{\ell_1 } c_{j , \sigma}^\dagger c_{j , \sigma} 
 + \Delta_1 \sum_{ j = 1}^{ \ell_1  } \{ c_{j , \uparrow} c_{ j , \downarrow } + 
 c_{ j , \downarrow }^\dagger 
 c_{j , \uparrow}^\dagger \}
 \nonumber \\
 &  - & w_2 \sum_\sigma \sum_{ j = 1}^{ \ell_2 - 1 } \{ d_{ j , \sigma}^\dagger d_{ j + 1 ,
 \sigma} + 
 d_{ j + 1 , \sigma }^\dagger d_{j , \sigma} \} - \mu_2 \sum_\sigma
 \sum_{ j = 1}^{\ell_2 } d_{j , \sigma}^\dagger d_{j , \sigma} 
 + \Delta_2 \sum_{ j = 1}^{ \ell_2 } \{ d_{j , \uparrow} d_{ j  , \downarrow} 
 + d_{ j , \downarrow}^\dagger   d_{j , \uparrow}^\dagger \}
 \nonumber \\
 &  - &\sum_\sigma\tau \{ [ c_{1 , \sigma}^\dagger d_{ \ell_2 , \sigma} + 
 d_{1 , \sigma}^\dagger c_{ \ell_1 , \sigma } ] e^{ \frac{i}{4} \Phi } 
 + [ d_{ \ell_2 , \sigma}^\dagger c_{1 , \sigma} + 
 c_{ \ell_1 , \sigma}^\dagger d_{1 , \sigma} ] e^{  - \frac{i}{4} \Phi }  \}
 \:\:\:\: . 
 \label{sw.1}
\end{eqnarray}
\noindent
Going through the same steps as in the p-wave case, we eventually find that 
the transfer matrix of the ring is now given by

\beq  
{\cal M}_{ s - wave } [ E ; \Phi ; \ell_1 ; \ell_2 ]  =  [ \tilde{\Upsilon}^{(2)}  ( E ) ]^{-1} 
\cdot  \tilde{\Sigma}_2^{-1}  [ \Phi] \cdot \tilde{\Omega}_1 [ \Phi ]  \cdot [ 
\tilde{\Upsilon}^{(1)}  ( E ) ] 
\cdot {\bf T}_1 ( E  ; \ell_1)  
  \cdot [ \tilde{\Upsilon}^{(1)}  ( E ) ]^{-1} \cdot \tilde{\Sigma}_1^{-1} [ \Phi] \cdot 
 \tilde{\Omega}_2 [ \Phi ]  \cdot [ \tilde{\Upsilon}^{(2)}  ( E ) ] \cdot {\bf T}_2 ( E  ; \ell_2)  
 \:\:\:\: . 
 \label{sw.13}
 \eneq
 \noindent
The $\tilde{\Sigma}_a  [ \Phi] ,  \tilde{\Omega}_a [ \Phi ]$    matrices in Eq.(\ref{sw.13}) 
are simply obtained by setting $\Delta_a$ to 0 respectively in Eq.(\ref{pw.16}) and 
in Eq.(\ref{pw.18}). The matrix  $[ \tilde{\Upsilon}^{(a)}  ( E ) ] $ is given by

\beq
[ \tilde{\Upsilon}^{(a)}  ( E ) ] = \left[ \begin{array}{cccc}
e^{ i k_p^{(a)} } u_p^{(a)} &    e^{ - i k_p^{(a)} } u_p^{(a)}     &
e^{ -  i k_h^{(a)} } u_h^{(a)} &    e^{  i k_h^{(a)} } u_h^{(a)} \\
e^{ i k_p^{(a)} } v_p^{(a)} &     e^{ - i k_p^{(a)} } v_p^{(a)}     &
e^{ -  i k_h^{(a)} } v_h^{(a)} &     e^{  i k_h^{(a)} } v_h^{(a)} \\
  u_p^{(a)} &      u_p^{(a)}     &
 u_h^{(a)} &     u_h^{(a)} \\
  v_p^{(a)} &      v_p^{(a)}     &
  v_h^{(a)} &     v_h^{(a)} 
                                   \end{array} \right]
\:\:\:\: ,
\label{sw.11}
\eneq
\noindent
with   $u^{(a)} , v^{(a)}$ determined as nontrivial solutions of 
the algebraic system 

\begin{eqnarray}
 0 &=& \{ E + 2 w_a \cos ( k_a ) + \mu_a \} u^{(a)} - \Delta_a v^{(a)} \nonumber \\
 0 &=& - \Delta_a u^{(a)}  + \{ E - 2 w_a \cos ( k_a ) - \mu_a \} v^{(a)}
 \:\:\:\: , 
 \label{sw.5}
\end{eqnarray}
\noindent
for  $k_1 , k_2$ solving the secular equation

\beq
E^2 = [ 2 w_1 \cos (k_1 ) + \mu_1 ]^2 + \Delta_1^2 = 
 [ 2 w_2  \cos (k_2 ) + \mu_2 ]^2 + \Delta_2^2
 \:\:\:\: , 
 \label{sw.6}
 \eneq
 \noindent
 and $k_p^{(a)} , k_h^{(a)}$ defined by 
 
 \begin{eqnarray}
  \cos [ k_p^{(a)} ] &=& - \frac{\mu_a}{2 w_a} - \sqrt{\frac{E^2 - \Delta_a^2}{4 w_a^2}}
  \nonumber \\
  \cos [ k_h^{(a)} ] &=& - \frac{\mu_a}{2 w_a}+ \sqrt{\frac{E^2 - \Delta_a^2}{4 w_a^2}}
  \:\:\:\: . 
  \label{sw.7}
 \end{eqnarray}
\noindent
Besides the differences in the form of the matrices appearing in Eqs.(\ref{pw.21},\ref{sw.13}), an important 
point   to stress  is that both matrices are  block-factorizable, with the factorization corresponding to the possibility of 
regarding the system as a sequence of homogeneous regions separated by interfaces. Indeed, 
the TM for a one-dimensional system comes out to be simply the ordered product of the 
matrices corresponding to the homogeneous regions and of the ones corresponding to 
the interfaces, taken in the appropriate sequence. From this respect, the matrices 
corresponding to each homogeneous region and to each interface are sort of ''building blocks''
of the global transfer matrix (see Fig.\ref{tm} for a sketch of the factorizability of 
the matrix).

\noindent

\section{Calculation of the persistent current} 
 \label{calc_persist}

In this section we compute the persistent current in a number of  cases of 
interest.   To highlight the feature related to the emergence of MFs at the 
SN-interfaces,  it is worth comparing the results obtained 
in p-wave systems with the ones obtained in s-wave systems.   Therefore, in the following 
we perform the calculation in both cases, by using  
Eq.(\ref{ra.11}), with the transfer matrix computed according to Eq.(\ref{pw.21})
(p-wave case), or to Eq.(\ref{sw.13}) (s-wave case). To keep in touch with 
the results of Ref.[\onlinecite{pientka}], we begin with the calculation of the 
current in the case of a superconducting ring interrupted by a weak link though, at
variance with the discussion of Ref.[\onlinecite{pientka}], we will not assume fermion parity
conservation. As stated above,  
for comparison, we also compute the current in the case of a ring made with 
an s-wave superconductor.

\subsection{Persistent current across a superconducting ring interrupted by 
a weak link}
\label{weali.1}

A p-wave superconducting ring interrupted by a weak link can be physically
realized at a semiconducting quantum wire with a 
sizeable spin-orbit coupling (e.g., an InAs wire) deposited onto
a bulk superconducting ring pierced by a magnetic flux $\Phi$.
The combined effect of spin-orbit coupling, Zeeman spin splitting and proximity-induced
superconductivity from the bulk superconductor underneath has been shown to make the 
wire effectively behave as a one-dimensional p-wave superconductor \cite{Lutchyn,oreg}.  As for 
what concerns a concrete proposal of an experimental realization of the p-wave HR,  
we refer to Refs.[\onlinecite{pientka,buttiker}]. Specifically, we assume that 
the weak link is actually realized as a physical interruption of the superconducting
ring with a tiny insulating layer, which cuts the current within the 
superconductor, thus allowing the persistent current to only flow across the semiconducting
nanowire. In fact, among other advantages, this geometry allows for recovering as only
detectable current the one flowing through the semiconducting wire, which is what we 
are eventually interested in.  
An important point to stress is that it is typically difficult to keep the ring perfectly 
isolated from the substrate, so to avoid fermion parity non-conserving relaxation processes 
[29,45]. Moreover, in order for the  grand-canonical like description of 
our system we employ here to be reliable, one has to think of a ring in contact with a substrate,
working as an electronic reservoir. Therefore, we should not  expect fermion parity to be preserved
here. For this reason, though one  might in principle account for fermion parity conservation  by
implementing some pertinent adapted version of the approach presented in 
Ref.[37], throughout all our paper we assume fermion parity not to be conserved which, 
nevertheless, does not affect the possibility of probing emergent MFs by means of an appropriate 
persistent current measurement. 

\begin{figure}
\includegraphics*[width=.75\linewidth]{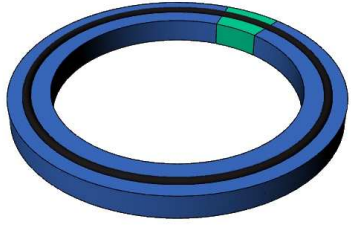}
\caption{Sketch of the superconducting ring interrupted by a weak link as 
introduced and discussed in Ref.[\onlinecite{pientka}]. A semiconducting ring (depicted as 
a solid black line) is deposited on top of a bulk superconducting ring, interrupted by 
a tiny insulating layer (light blue sector). The superconductor induces superconductivity
within the semiconducting wire by proximity effect. All the current circulating across the system
is due to tunneling effect within the  semiconducting wire.} \label{weaklink}
\end{figure}

In Fig.\ref{weaklink} we provide a sketch of the system we 
are considering here, that is, a homogeneous ring interrupted by a weak link. 
The corresponding model Hamiltonian  can be recovered from  Eq.(\ref{pw.1})
in the p-wave case and from Eq.(\ref{sw.1}) in the s-wave case, by setting to zero the 
length of one of the two regions. In the former case, it is 
given by 

\begin{figure}
\includegraphics*[width=.65\linewidth]{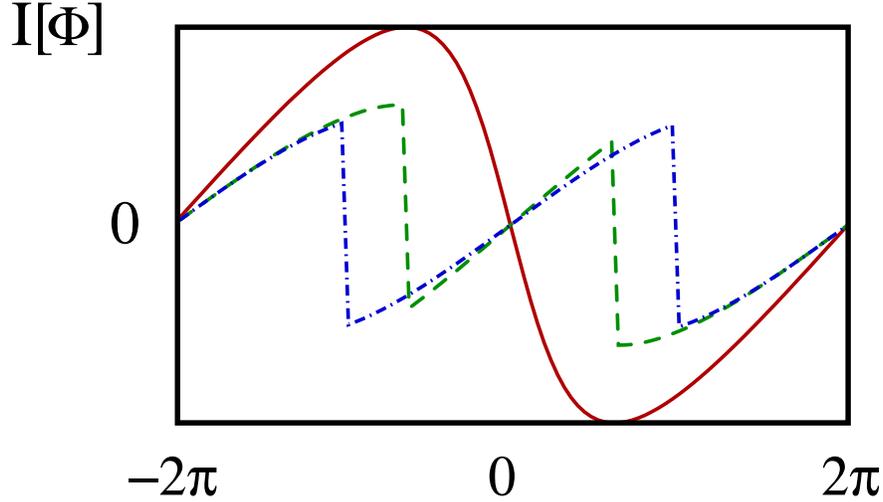}
\caption{Plot of the persistent current $I [ \Phi ]$ {\it vs.} $\Phi$ (in units
of $\Phi_0$) for the  p-wave mesoscopic ring with a weak link
described by Eq.(\ref{wl.1}). The parameters are chosen so that $\mu = 0$, 
$\Delta = 0.2$ and $\tau = 0.5$ (in units of $w$), 
$\ell = 4$ (full red curve), $\ell = 8$ (dashed green curve), $\ell = 40$ (dot-dashed
blue curve). The estimated Majorana hybridization length in this case 
 is $\ell_M \approx 5$.} \label{plot_mu_zero}
\end{figure}

\noindent

\begin{eqnarray}
 H_{p-w} &=& - w \sum_{ j = 1}^{ \ell - 1 } \{ c_j^\dagger c_{ j + 1 } + 
 c_{ j + 1}^\dagger c_j \} - \mu \sum_{ j = 1}^{\ell  } c_j^\dagger c_j 
 + \Delta \sum_{ j = 1}^{ \ell  - 1 } \{ c_j c_{ j + 1 } + c_{ j + 1}^\dagger 
 c_j^\dagger \} 
 \nonumber \\
 &  - &\tau \{   c_1^\dagger c_{ \ell}   e^{ \frac{i}{2}  \Phi } 
 + c_{ \ell}^\dagger c_1   e^{  - \frac{i}{2} \Phi }  \}
 \:\:\:\: , 
 \label{wl.1}
\end{eqnarray}
\noindent
while in the latter case it can be presented as 

\begin{eqnarray}
 H_{s-w} &=& - w \sum_\sigma \sum_{ j = 1}^{ \ell - 1 } \{ c_{ j , \sigma}^\dagger 
 c_{ j + 1  , \sigma } +   c_{ j + 1 , \sigma }^\dagger c_{j , \sigma} \} - 
 \mu \sum_\sigma \sum_{ j = 1}^{\ell_1 } c_{j , \sigma}^\dagger c_{j , \sigma} 
 + \Delta \sum_{ j = 1}^{ \ell  } \{ c_{j , \uparrow} c_{ j , \downarrow } + 
 c_{ j , \downarrow }^\dagger 
 c_{j , \uparrow}^\dagger \}
 \nonumber \\ 
 &  - &\sum_\sigma \tau \{   c_{1 , \sigma}^\dagger c_{ \ell  , \sigma} e^{ \frac{i}{2} \Phi } 
 +  c_{ \ell  , \sigma}^\dagger c_{1 , \sigma}  e^{  - \frac{i}{2} \Phi }  \}
 \:\:\:\: . 
 \label{wl.2}
\end{eqnarray}
\noindent
Let us note that, in this case, the flux is fully ''loaded'' on the single hopping term.
The transfer matrix derived from Eqs.(\ref{wl.1},\ref{wl.2}) 
can therefore be simply expressed in terms of the ones we provide
in section \ref{mswave} by simply setting one of the two lengths (say $\ell_2$) 
to 0. We therefore obtain 

 \beq
 {\cal M}_{ p-w} [ E ; \ell ; \Phi ] = [ \Upsilon ( E ) ]^{-1} \cdot \Sigma^{-1} [ 2 \Phi ] 
 \cdot \Omega [ 2 \Phi ] \cdot [ \Upsilon ( E ) ] \cdot {\bf T} [ E ; \ell ] 
 \:\:\:\: , 
 \label{wl.3}
 \eneq
 \noindent
 and, similarly

 \beq
 {\cal M}_{ s-w} [ E ; \ell ; \Phi ] = [\tilde{\Upsilon} ( E ) ]^{-1} \cdot 
 \tilde{\Sigma}^{-1} [ 2 \Phi ] 
 \cdot \tilde{\Omega} [ 2 \Phi ] \cdot [ \tilde{\Upsilon} ( E ) ] \cdot {\bf T} [ E ; \ell ] 
 \:\:\:\: . 
 \label{wl.4}
 \eneq
 \noindent
To compute the current,  let us start with the   ring  described by the TM in Eq.(\ref{wl.3}). In Fig.\ref{plot_mu_zero}, 
we plot $I [ \Phi ]$ {\it vs.}  $\Phi$ 
for the values of the parameters reported in the caption, particularly for 
a chemical potential $\mu=0$. At zero chemical potential, the p-wave superconductor lies well 
within the topological region, with two MFs $\gamma_L , \gamma_R$ 
localized at its endpoints. This allows us to provide a simple interpretation of the 
curves  we draw in Fig.\ref{plot_mu_zero} 
at different values of the length $\ell$ of the superconductor. The key parameter here is 
the ''hybridization length''  $\ell_M$ between $\gamma_L$ and $\gamma_R$, which we estimate 
according to the derivation of  the Appendix \ref{exw}. At $\mu = 0$ and 
for the values of the parameters we used, from Eq.(\ref{mfo.19}) we obtain $\ell_M \sim 5$.
Therefore, when considering the largest ring ($\ell = 40$), we may safely neglect 
the overlap between $\gamma_L $ and $\gamma_R$ across the superconducting region
and accordingly describe the low-energy
excitations of the system by approximating 
the fermion operators in the tunneling term of the total Hamiltonian [second row of 
Eq.(\ref{wl.1})] by means of the truncated mode expansion in Eq.(\ref{mfo.26}). As 
a result, we obtain  the effective low-energy Hamiltonian $H_{p ; Eff}$ for the ring, 
given by

\beq
H_{p ; Eff} \approx - \epsilon_0 \: \cos \left( \frac{\Phi}{2} \right) 
\left\{ 2 \Gamma^\dagger \Gamma  - 1 \right\} 
\:\:\:\: , 
\label{wl.5}
\eneq
\noindent
with $\epsilon_0 \propto \tau$ and the Dirac fermion operator $\Gamma$ related 
to $\gamma_L , \gamma_R$ by means of Eqs.(\ref{mfo.32}). We now use Eq.(\ref{wl.5}) as
a main reference to discuss the behavior of the current for large-$\ell$. In fact, while 
one should, in principle, consider the contributions arising from all the populated single-quasiparticle
states at any energy (which is exactly done in the calculation we performed based on 
our TM approach), based upon arguments similar to the ones provided in Refs.[\onlinecite{giu_af1,giu_af2}],
in the large-$\ell$ limit we expect that the result for $I [ \Phi ]$ can be safely 
recovered by taking into account only low-energy states of the system. 
Now,  for  $ - \pi  < \Phi < \pi$, Eq.(\ref{wl.5}) tells us that 
the ground state has the $\Gamma$-level populated. As $\Phi$ crosses $\pi$, 
there is a  crossing between the filled- and the empty-$\Gamma$ 
state which, in the absence of constraints on fermion parity conservation, 
makes the system ''jump'' from the populated to the empty $\Gamma$-fermion 
state, with the corresponding jump in the current evidenced at $\Phi = \pi$ in 
the blue curve of  Fig.\ref{plot_mu_zero}, corresponding to $\ell = 40$. By symmetry $\Phi \to - \Phi$,
an analogous jump is observed at $\Phi = - \pi$. The total current is periodic, with 
period equal to $2 \pi$, due to the sequential level crossings at 
$\Phi = 2 \pi k + \pi$, with integer $k$ \cite{oreg}.

\begin{figure}
\includegraphics*[width=.8\linewidth]{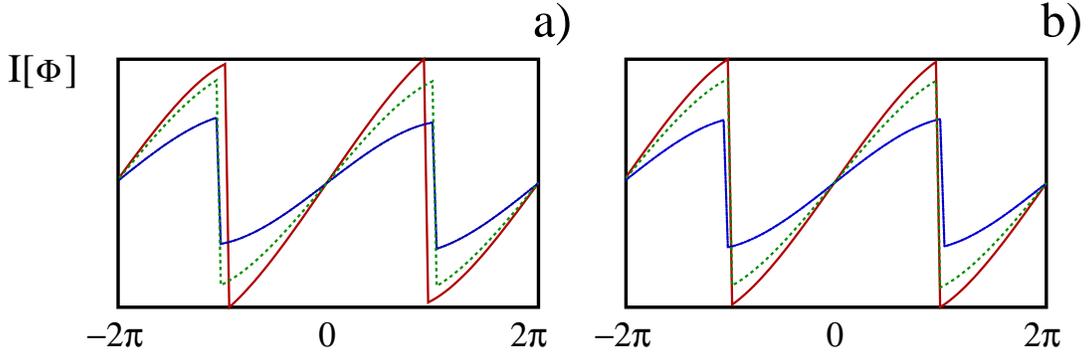}
\caption{Plot of the persistent current $I [ \Phi ]$ {\it vs.} $\Phi$ (in units
of $\Phi_0$) for the  p-wave mesoscopic ring with a weak link
described by Eq.(\ref{wl.1}). The parameters are chosen so that   
$\Delta = 0.2$ and $\tau = 0.5$ (in units of $w$). The curves are plotted for 
various values of $\mu$ at fixed $\ell$. Specifically: \\  
{\bf a)} $\ell = 16$, $\mu=0.0$ (full red curve), $\mu=0.9$ (dashed green curve),
$\mu=1.5$ (dot-dashed blue curve); \\
{\bf b)} $\ell = 40$, $\mu=0.0$ (full red curve), $\mu=0.9$ (dashed green curve),
$\mu=1.5$ (dot-dashed blue curve).} \label{varying_mu}
\end{figure}

\noindent
At variance, as $\ell = 4$, the hybridization between the MFs
across the topological superconductor is no longer negligible. As a result, 
at low energy the system is described by an effective Hamiltonian such as 
the one in Eq.(\ref{mfo.32a}), with a modulation with $\Phi$ of the energy splitting between
the empty- and the filled-state, which never
closes (avoided level crossing). In this case, the persistent current is only
supported by Cooper pair tunneling across the weak link, which restores 
a $4 \pi$-periodicity in $\Phi$  \cite{oreg}. 
Again, this is consistent with our plot in  Fig.\ref{plot_mu_zero} for $\ell = 4$, 
with the intermediate case $\ell = 8$ lying in between the two ''asymptotic'' cases. 
As a further check, we report in Fig.\ref{varying_mu} the plots of $I [ \Phi ]$ generated
by keeping $w=1 , \Delta = 0.2 , \tau= 0.5$ and varying $\mu$, with $\ell = 16$ 
(Fig.\ref{varying_mu} {\bf a}) and $\ell = 40 $ (Fig.\ref{varying_mu} {\bf b}) 
(note that  $\ell_M \sim 5$, as estimated above). From 
the plots we draw for $\mu = 0.0, 0.9 , 1.5$, we see that, 
increasing $\mu$ toward the critical value $\mu = 2$ at which the topological 
phase transition takes place \cite{kita}, effectively corresponds to 
increasing $\ell_M$. This is expected from the results of the Appendix \ref{exw}, where
we show that the hybridization between $\gamma_L$ and $\gamma_R$ scales 
as $e^{ - \ell / \ell_M }$. Thus, again our results appear to be consistent 
with the low-energy dynamics of our system as inferred from appendix \ref{exw} 
and from the discussion reported in Ref.[\onlinecite{oreg}].

\begin{figure}
\includegraphics*[width=.7\linewidth]{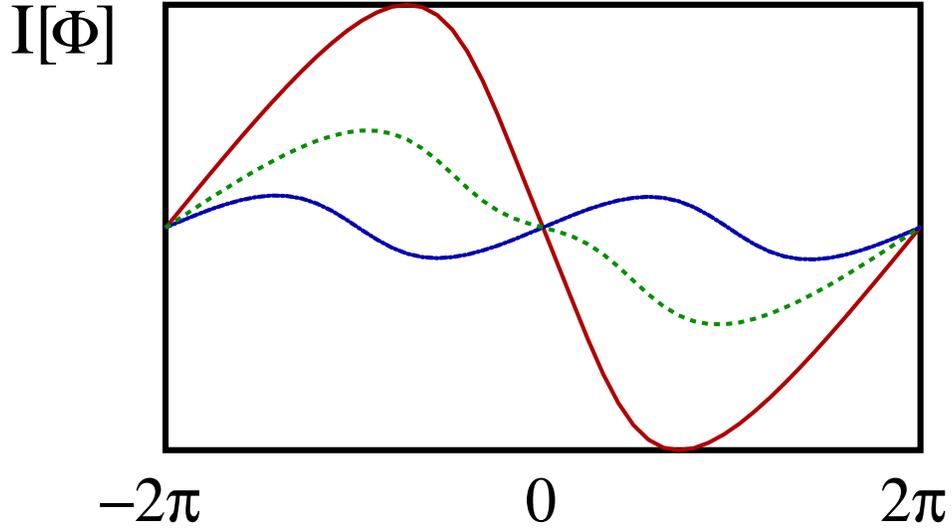}
\caption{Plot of the persistent current $I [ \Phi ]$ {\it vs.} $\Phi$ (in units
of $\Phi_0$) for the  s-wave mesoscopic ring with a weak link
described by Eq.(\ref{wl.2}). The parameters are chosen so that   
$\Delta = 0.2$ and $\tau = 0.5$ (in units of $w$). The curves are plotted for 
various values of $\ell$ at  $\mu = 0$. Specifically:  
$\ell = 4$ (full red curve), $\ell = 16$ (dashed green curve),
$\ell = 40$ (dot-dashed blue curve). The crossover from a $4 \pi$ periodicity 
for $\ell < \xi_0 \sim 10$ (see text) to a $2 \pi$-periodicity for $\ell > \xi_0$ is apparent.
} \label{swave}
\end{figure}

\noindent
By contrast, we now discuss the current across  an s-wave ring. Despite the lack of 
low-energy MFs in such a system, the crossover in the 
periodicity of $I[ \Phi ] $ from $4 \pi$ to $2 \pi$ is known to take place 
as the length $\ell_S$ of the superconducting region crosses over from values 
lower than the coherence length $\xi_0$ to values higher than $\xi_0$ \cite{butti_1}. 
Such a crossover corresponds to a crossover in the ''physical nature'' of 
$I [ \Phi ]$: from a $4 \pi$-persistent current in a mesoscopic, effectively 
normal, system to a $2 \pi$-periodic current, analogous to the Josephson 
supercurrent in an SNS-junction \cite{butti_1,montam}. In Fig.\ref{swave}, we plot
the exact results for $I [ \Phi ]$ obtained using our TM-approach for 
the system parameters $w = 1 , \Delta = 0.2 , \mu = 0 , \tau = 0.5$ and 
for various values of $\ell$. 
In our specific case, having as model Hamiltonian the one in  Eq.(\ref{wl.2}), as we are setting to 1 the lattice step, 
an acceptable estimate for $\xi_0$ is  $\xi_0 \sim 2 w / \Delta$. For the numerical 
values of the parameters we chose to generate Fig.\ref{swave}, this implies 
$\xi_0 \sim 10$. Such an estimate is definitely consistent with our results: on increasing 
$\ell$ from $\ell = 4$ to $\ell = 40$, we ultimately see a crossover in the periodicity of $ I [ \Phi ]$ definitely 
similar to the one we found for the p-wave superconducting ring with a weak link, though
without the jumps in the current due to the $\Gamma$-fermion level crossings. To conclude this section, 
let us stress once more that our technique does provide us with the exact result for $I [ \Phi ]$ at
a generic value of the system parameters, whether the superconductor is p-wave, or s-wave, etc. To recover
the final result one just needs to construct the appropriate TM and to 
numerically compute a single integral for various values of $\Phi$. Using the standard
approach, based on the solution of the secular equations for the allowed values of 
the momenta at fixed $\Phi$, and  eventually taking the derivative with respect to $\Phi$ to obtain the current is, in 
general, much less straightforward and, typically, exact results cannot be provided and 
different approximations must be implemented to attack different regimes such as the 
short-ring, or the long-ring limit (see  Ref.[\onlinecite{montam}] for a careful and 
valuable discussion about this point). At variance, as we are showing here, our approach 
applies to any specific case, with potentially no limitations at all.   It allows us to recover the full  
plots of the persistent current at various system scales, which we show in this section and in the following one: an original result that complements and 
extends the analysis of Ref.[\onlinecite{pientka}], where the analysis of the current was performed by just considering how the 
relative weight of the first two harmonics (in $\Phi$) varies, as a function of the system parameters.  
In the following, to discuss a 
further application of our technique, we  consider a hybrid ring, made by a 
p-wave superconducting segment of length $\ell_S$ and a normal segment of length 
$\ell_N$: this can be regarded as a generalization of the ring with a weak link  which, as we are going to discuss, 
opens the way to a number of interesting physical effects.

\noindent

\subsection{Persistent current across a hybrid N-S ring}
\label{hybri.NS}

\begin{figure}
\includegraphics*[width=.75\linewidth]{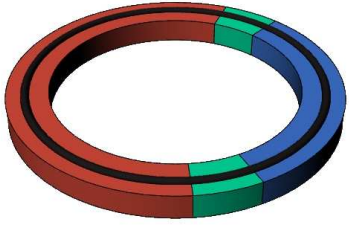}
\caption{Sketch of the  hybrid N-S ring with the two regions (red and blue, respectively) 
separated by two a weak links. A possible practical realization of such a system is the same
as discussed before for the superconducting ring interrupted by a weak link.} \label{tworegion}
\end{figure}

\begin{figure}
\includegraphics*[width=.55\linewidth]{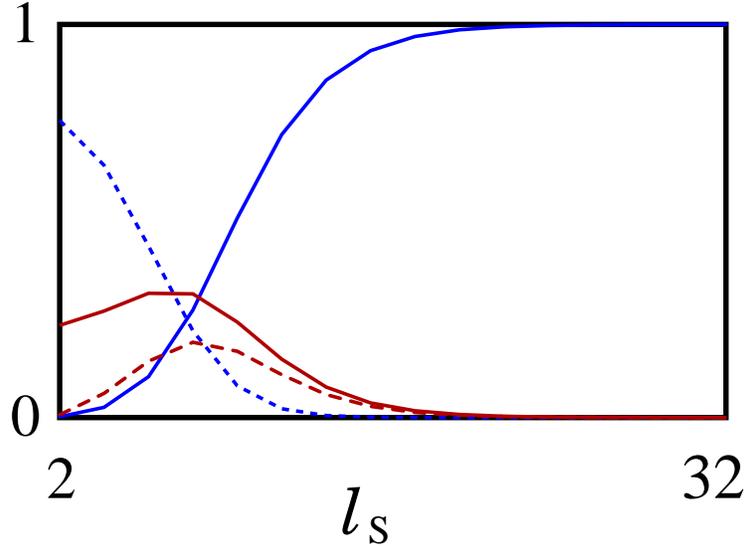}
\caption{Plot of the normal and Andreev reflection coefficients at 
the SN-interface and of the normal transmission and CAR coefficients 
for the p-wave superconductor-normal hybrid ring discussed in section 
\ref{hybri.NS} as a function of $\ell_S$. To generate the plot we have chosen the 
system's parameters so that $w_1 = w_2 = 1$, $\Delta = 0.2$, $\tau = 0.5$, which corresponds 
to $\ell_M \approx 6$. The various curves correspond to the coefficients as \\
- Full blue curve: Andreev reflection coefficient at the SN-interface; \\
- Dashed blue curve: Normal reflection coefficient at the SN-interface;\\
- Full red curve: Normal transmission across the superconducting region;\\
- Dashed red curve: Crossed Andreev reflection across the superconducting region. \\
Apparently, for $\ell_S > \ell_M$ all the processes are suppressed, except the 
Andreev reflection at the SN-interface, with the corresponding coefficient saturating 
to 1.
} \label{coefficients}
\end{figure}

In this section we discuss the persistent current across a hybrid ring, composed
of a superconducting segment of length $\ell_S$ and of a normal segment of length 
$\ell_N$. Such a system can be regarded as a generalization of the 
ring interrupted by a weak link discussed in Ref.[\onlinecite{pientka}], in which 
one induced superconductivity by proximity only in a part of the ring, leaving a 
finite normal region of length $\ell_N$. In Fig.\ref{tworegion} we provide a 
sketch of the system we discuss here. Again, for comparison, we consider both cases of a p-wave and of an s-wave 
superconducting region.  The corresponding model Hamiltonian can then be   obtained from $H$ in Eq.(\ref{pw.1}) by setting $\ell_1 \to \ell_S$, 
$\ell_2 \to \ell_N$, $\Delta_1 \to \Delta , \Delta_2 = 0$ and from $H$ in Eq.(\ref{sw.1}), taken in 
the same limit. To spell out the behavior of $ I [ \Phi ]$ in the various 
regimes of interest, let us first focus onto the p-wave case. Specifically, we 
compute $ I [ \Phi ]$ at given $\tau$ and $\Delta$ for $w_1 = w_2 \equiv w$ and for 
various values of $\ell_S = \ell_N \equiv \ell$. To further simplify the calculation 
we restrict  ourselves to the particle-hole symmetric case, $\mu = 0$.
\noindent
\begin{figure}
\includegraphics*[width=.97\linewidth]{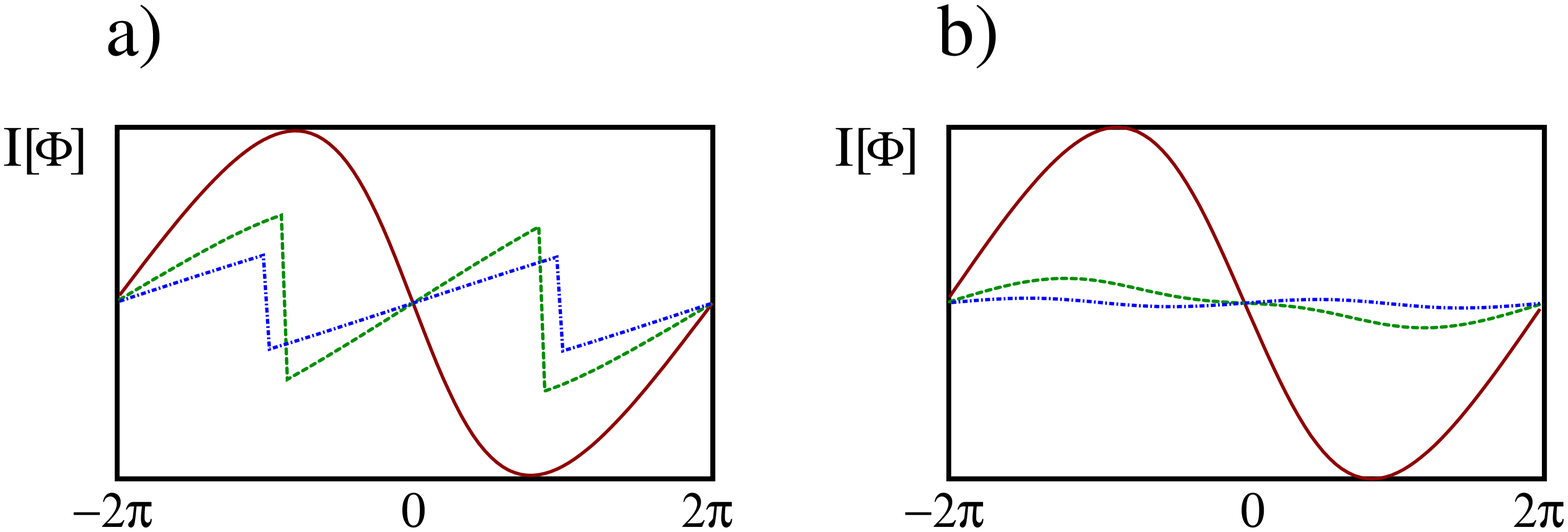}
\caption{Plot of $ I[ \Phi ]$ {\it vs.} $\Phi$ for a hybrid normal - superconducting 
ring. \\
 {\bf a)} Plot of $ I[ \Phi ]$ {\it vs.} $\Phi$ for a hybrid normal-p-wave superconducting
ring for $w_1 = w_2 = 1$, $\Delta = 0.2$, $\tau = 0.5$, $\mu = 0$, which corresponds 
to $\ell_M \approx 6$, $\xi_{ K , M} \approx 16$, for $\ell_S = \ell_N = 4$ (solid red
curve), $\ell_S = \ell_N = 16$ (dashed green curve), $\ell_S = \ell_N = 40$ 
(dot-dashed blue curve). The crossover to a sawtooth behavior is evident 
for $\ell = 40$; \\
{\bf b)}. Plot of $ I[ \Phi ]$ {\it vs.} $\Phi$ for a hybrid normal-s-wave superconducting
ring for $w_1 = w_2 = 1$, $\Delta = 0.2$, $\tau = 0.5$, for $\ell_S = \ell_N = 4$ (solid red
curve), $\ell_S = \ell_N = 16$ (dashed green curve), $\ell_S = \ell_N = 40$ 
(dot-dashed blue curve). There is no  crossover in the functional form of $ I [ \Phi ]$, but
a mere scaling of $I [ \Phi ]$ that is $\sim \ell_N^{-1}$.
} \label{all_wave_finite}
\end{figure}
\noindent
In Fig.\ref{all_wave_finite}{\bf a)}, 
we plot $ I [ \Phi ]$ {\it vs.} $ \Phi$ for $\ell = 4 , 16 , 40$, with the values for the 
system's parameters chosen as in the caption. The behavior of $ I [ \Phi ]$ 
depends on the system  size in relation to the length scales  determined by the system parameters. 
At $\mu = 0$ the p-wave superconductor lies
within its topological phase, with corresponding localized MFs emerging at its endpoints. 
Taking again $\ell_M$ as a reference length scale, when $\ell \leq \ell_M$, the two 
MFs are hybridized into a Dirac mode $\Gamma$. As a result, the MFs have no significant effects 
on the current, which is $4 \pi$-periodic, consistently with the expected behavior of the system 
as a mesoscopic normal ring \cite{oreg}. On increasing $\ell$, when 
$\ell > \ell_M$, the hybridization between the MFs becomes negligible and, 
accordingly, in the absence of fermion parity conservation, $ I [ \Phi ]$ becomes 
$2 \pi$-periodic, with jumps at $\Phi = \pi + 2 \pi k$. In addition to the periodicity, also the shape of
$ I [ \Phi ]$ depends on $\ell$. This is due to the Kondo-like hybridization between
the MFs and the excitation modes within the normal region of the ring, 
which takes place when $\ell \geq \xi_{ K , M }$, with the Kondo-Majorana hybridization (KMH) 
length $\xi_{ K , M } \sim [ (2 w)^2 / \tau^2] $ \: \cite{giuaf_3}. At the onset of KMH, $ I [ \Phi ]$ is expected to cross from 
a discontinuous sinusoidal dependence on $\Phi$ to a sawtooth-like shape \cite{giuaf_3}. Physically, this can be 
understood by recalling that, as $\ell$ becomes large, the systematic cancellation of contributions
from high-energy states makes only low-energy states in the normal region next to the Fermi level contribute
to $ I [ \Phi ]$. The physical processes at the SN-interfaces that determine these states can be 
inferred from Fig.\ref{coefficients}, where, as a function of $\ell_S$, we plot the scattering coefficients 
across the superconducting regions corresponding to normal reflection at the SN-interfaces and to 
normal transmission across the superconducting region, as well as the coefficients corresponding to 
Andreev reflection (AR) at the interfaces and to ''crossed Andreev reflection'' (CAR) across the superconducting
region \cite{law_lee,choi}.  As it clearly appears from Fig.\ref{coefficients}, as soon as $\ell_S > \ell_M$, all the 
coefficients drop to 0 but the one corresponding to AR, which saturates to 1. This evidences that, as 
$\ell > \ell_S$, AR is the only physical process that takes place at low energy, which implies
the sawtooth behavior in $ I [ \Phi ] $ evidenced in Fig.\ref{all_wave_finite}{\bf a)}. By comparison, 
in Fig.\ref{all_wave_finite}{\bf b)}, we plot $ I [ \Phi ]$ {\it vs.} $\Phi$ for the 
same values of the various parameters as in  Fig.\ref{all_wave_finite}{\bf a)}, but for 
an s-wave superconductor. Here we see that, on increasing $\ell$, the current 
still shows the crossover from a $4 \pi$-periodic curve to a $2 \pi$-periodic curve, but that 
the absence of low-energy Majorana modes eventually hybridizing with the modes in 
the normal region as $\ell \geq \xi_{ K , M }$ yields no crossover in the functional 
form of $ I [ \Phi ]$ from a sinusoidal to a sawtooth behavior. The only relevant additional
feature that takes place on varying $\ell$ is, indeed, the expected scaling of 
$ I [ \Phi ]$ $ \sim \ell^{-1}$ [see discussion in the next section]. Thus, the crossover in 
the functional form of $ I [ \Phi]$ can actually be regarded as a direct evidence for 
the existence of low-energy Majorana modes at the endpoints of the topological 
superconductor in the p-wave hybrid ring. 

\begin{figure}
\includegraphics*[width=.95\linewidth]{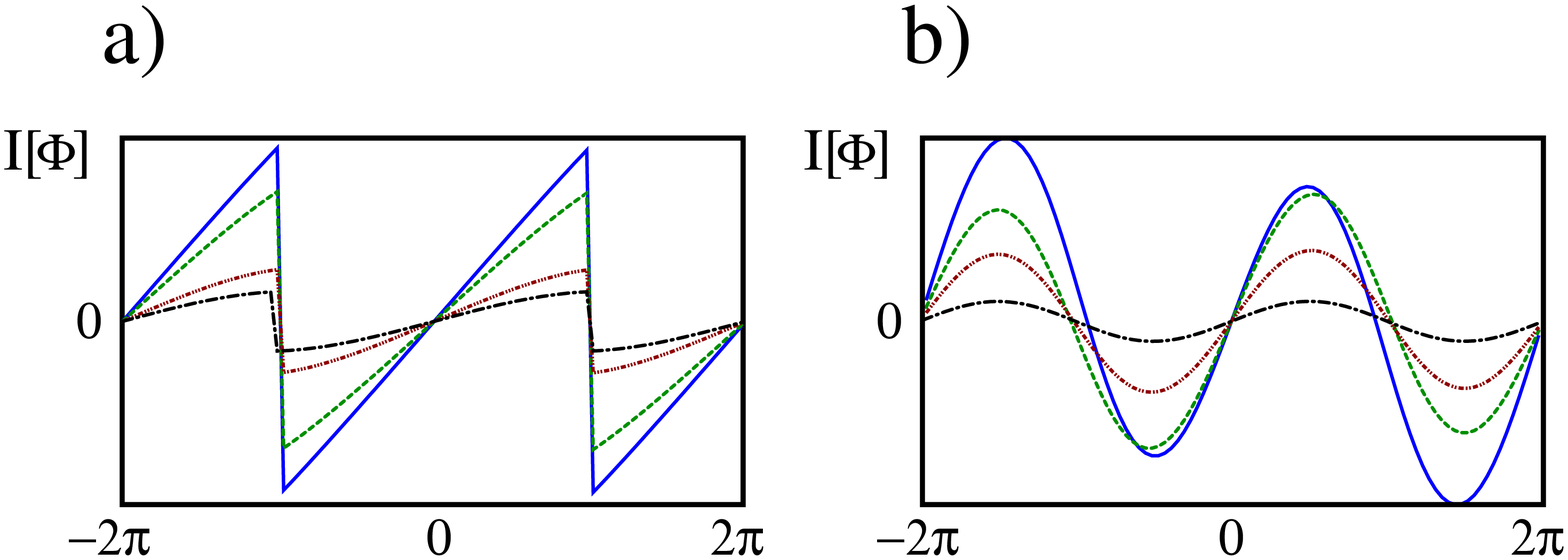}
\caption{Plot of $ I[ \Phi ]$ {\it vs.} $\Phi$ for a hybrid normal - superconducting 
ring. \\
 {\bf a)} Plot of $ I[ \Phi ]$ {\it vs.} $\Phi$ for a hybrid normal-p-wave superconducting
ring for $w_1 = w_2 = 1$, $\Delta = 0.2$, $\tau = 0.5$, $\mu_N = 0$, which corresponds 
$\xi_{ K , M} \approx 16$,  $\ell_S = \ell_N = 40$ and \\
- $\mu_S / w = 0.0$ (solid blue curve)\\
- $\mu_S / w = 1.5$  (dashed green curve) \\
- $\mu_S / w = 1.9$  (dot-dashed red curve) \\
- $\mu_S / w = 1.95$  (dot-dot-dashed black curve) \\
 {\bf b)} Same as in {\bf a)}, but for   a hybrid normal-s-wave superconducting
ring for $w_1 = w_2 = 1$, $\Delta = 0.2$, $\tau = 0.5$, $\mu_N = 0$,  
$\ell_S = \ell_N = 40$ and \\
- $\mu_S / w = 0.0$ (solid blue curve)\\
- $\mu_S / w = 1.0$  (dashed green curve) \\
- $\mu_S / w = 1.5$  (dot-dashed red curve) \\
- $\mu_S / w = 1.9$  (dot-dot-dashed black curve) \\
Panel {\bf a)} shows  a clear  smoothing of the sawtooth dependence of $ I [ \Phi ]$ on
$\Phi$ towards a sinusoidal plot as $\mu_S / w$ increases towards the critical 
value $\mu_S / w = 2$, at which Majorana fermions 
disappear.
} \label{with_mu}
\end{figure}

To further substantiate the above picture, in Fig.\ref{with_mu}{\bf a)}  we again plot 
$I [ \Phi ]$ {\it vs.} $\Phi$ for the system parameters numerically set as 
discussed in the caption. In drawing  Fig.\ref{with_mu}, we hold $\ell_N = \ell_S$ fixed 
at 40 (with $\Delta$ and $w$ chosen so that $\ell_M \sim 6$), as well as the chemical
potential within the normal region $\mu_N = 0$. At variance, we vary the 
chemical potential within the superconducting region $\mu_S$ starting from $\mu_S = 0$ 
till $\mu_S / w \sim 1.95$ (after which the loss of numerical precision appears not to give 
us reliable results). As highlighted by Kitaev \cite{kita}, as $\mu_S / w = 2$ the 
p-wave superconductor undergoes a (topological) quantum phase transition, characterized by the 
disappearance (for $ \mu_S / w > 2$) of the localized MFs at the 
endpoints of the superconductor. On approaching the phase transition from 
within the topological region, the closer the system is to the quantum critical point, 
the larger is the effective $\ell_M$. While the actual numerical estimate of $\ell_M$ 
as a function of e.g. $\mu_S / w$ at fixed $\Delta$ can in principle
be provided from Eq.(\ref{appe.1.5}) of appendix \ref{exw}, here we just focus on the 
consistency of our exact results with the expectation one gets from the above discussion.
In fact, the estimated KMH-length for the system used to derive $ I [ \Phi ]$ in 
 Fig.\ref{with_mu}{\bf a)}  is $\xi_{ K , M } \approx 16$ [see the discussion in 
 the caption of Fig.\ref{all_wave_finite}{\bf a)}, which is drawn at the same
 values of $w $ and $\tau$]. In  Fig.\ref{with_mu}{\bf a)}  we plot $ I [ \Phi ]$ for $\ell_S = 
 \ell_N = 40$, we see full KMH in the normal region for $\mu_N = 0$, as 
 evidenced by the sawtooth behavior of the current and by the corresponding 
 $ 2 \pi$-periodicity in $\Phi$. On increasing $\mu_S$ towards the 
critical value corresponding to the topological quantum phase transition, the 
nonnegligible hybridization between the MFs across the superconducting 
region is expected to compete with KMH and eventually to suppress it
(in fact, this can be regarded as a ''Majorana analog'' of the competition between
Kondo effect and RKKY-interaction in the two-impurity Kondo model 
\cite{twoimpukondo_1,twoimpukondo_2,twoimpukondo_2b,twoimpukondo_3},
just as the KMH can be regarded as the Majorana analog of the onset of the 
Kondo cloud in a Kondo system \cite{giuaf_3}). Consistently with the expectation, we
see that, on increasing $\mu_S$, the sawtooth is smoothed (with a sizable 
reduction in the critical current) and clearly evolves back towards a restoration 
of the $ 4 \pi$-periodicity that characterizes the regime $\ell_S \leq \ell_M $ (see discussion
above).  For comparison, in Fig.\ref{with_mu}{\bf b)} we draw similar
plots generated in the s-wave case. No particular changes in the functional
form of the current appear, except the reduction in the value of the current at a 
given $\Phi$. In our view, this result does actually enforce the reliability of a 
persistent current measurement as a tool to detect the presence of MFs 
at the endpoints of a p-wave superconductor in the topological phase.

\noindent 
It is important to stress once more that, while our approach allowed for readily  
studying the additional consequences of KMH in the case of a hybrid ring with 
an extended normal region, due to the increasing complexity of the system, this 
would be hardly doable within an alternative approach, without resorting to some 
''ad hoc'' approximations and possibly washing out some relevant physical effects. 
This enforces once more the importance of having an exact analytical formula, that 
applies independently of the specific values of the system parameters.  
Besides the possibility of exactly computing the current in a number of different 
physical systems by simply evaluating the integral in Eq.(\ref{ra.11}) for different 
values of $\Phi$, our approach also provide a remarkable tool to write, for 
large enough systems, $I [ \Phi ]$ in a power series of the inverse system size which, 
as we are going to discuss in the following, greatly simplifies the various calculations, 
by even  providing explicit analytic results for the current, in some 
simple cases.

\noindent
 \section{The large-ring limit}
 \label{large}
 
As  the ring size goes large, one may recast the integral formula for $ I [ \Phi ]$, Eq.(\ref{ra.11}), 
in an expansion in inverse powers of the length that gets large. As we discuss in the following, this 
leads to   a number of simplifications in the calculation of the current, similar to 
the ones implemented in Refs.[\onlinecite{giu_af1,giu_af2}], even leading, in some cases, to
a closed-form analytic formula for $ I [ \Phi ]$ {\it vs.} $\Phi$ at
given system parameters. In the following, we  discuss a few examples of calculation
of the persistent current in the large-size limit, also showing how a number of 
known results can be easily recovered within our formalism, once the appropriate 
limit is taken.

 \subsection{The limit of long superconducting region}
 \label{long_super}
 
The limit of long superconducting region is defined by sending 
$\ell_2 \to \infty$ in the system described by the model 
Hamiltonian in Eq.(\ref{pw.1}) (p-wave case), or 
by the Hamiltonian in Eq.(\ref{sw.1}) (s-wave case), after 
setting $\Delta_1 = 0$, so that region-1 is normal, and by
keeping $\ell_1$ finite. In this limit one expects to recover the results
for a the Josephson current across an SNS-hybrid junction. To show that 
this is, in fact, the case, we start  by rewriting  
${\rm det} \{  {\cal M}_{ p - wave } [ E ; \Phi ; \ell_1 ; \ell_2 ] - {\bf I} \}$ 
as

\begin{eqnarray} 
&& {\rm det} \{  {\cal M}_{ p - wave } [ E ; \Phi ; \ell_1 ; \ell_2 ] - {\bf I} \}  = c \: {\rm det} \{  {\bf T}_2^{-1}  ( E  ; \ell_2) \nonumber \\
&& - [ \Upsilon^{(2)}  ( E ) ]^{-1} \cdot  \Sigma_2^{-1}  [ \Phi] \cdot \Omega_1 [ \Phi ]  \cdot [ \Upsilon^{(1)}  ( E ) ] 
\cdot {\bf T}_1 ( E  ; \ell_1)  
  \cdot [ \Upsilon^{(1)}  ( E ) ]^{-1} \cdot \Sigma_1^{-1} [ \Phi] \cdot 
 \Omega_2 [ \Phi ]  \cdot [ \Upsilon^{(2)}  ( E ) ]  
 \:\:\:\: , 
 \label{pw.21x}
 \end{eqnarray}
 \noindent
 with $c$ being an over-all factor independent of $\Phi$ and, similarly, by rewriting   
 ${\rm det} \{  {\cal M}_{ s - wave } [ E ; \Phi ; \ell_1 ; \ell_2 ] - {\bf I} \}$ 
as 
 
 \begin{eqnarray}
 && {\rm det} \{  {\cal M}_{ s - wave } [ E ; \Phi ; \ell_1 ; \ell_2 ]  - {\bf I} \} =  c' \: {\rm det} \{  {\bf T}_2^{-1}  ( E  ; \ell_2) \nonumber \\ 
  &&- [ \tilde{\Upsilon}^{(2)}  ( E ) ]^{-1} 
\cdot  \tilde{\Sigma}_2^{-1}  [ \Phi] \cdot \tilde{\Omega}_1 [ \Phi ]  \cdot [ 
\tilde{\Upsilon}^{(1)}  ( E ) ] 
\cdot {\bf T}_1 ( E  ; \ell_1)  
  \cdot [ \tilde{\Upsilon}^{(1)}  ( E ) ]^{-1} \cdot \tilde{\Sigma}_1^{-1} [ \Phi] \cdot 
 \tilde{\Omega}_2 [ \Phi ]  \cdot [ \tilde{\Upsilon}^{(2)}  ( E ) ]  \}
  \:\:\:\: ,
  \label{sw.1x}
 \end{eqnarray}
 with, again, $c'$ being a constant independent of $\Phi$. As a next
 step, we  define the matrix ${\bf M}_p$ as 
 
 \beq
 {\bf M}_p  = [ \Upsilon^{(2)}  ( E ) ]^{-1} \cdot  \Sigma_2^{-1}  [ \Phi] \cdot \Omega_1 [ \Phi ]  \cdot [ \Upsilon^{(1)}  ( E ) ] 
\cdot {\bf T}_1 ( E  ; \ell_1)  
  \cdot [ \Upsilon^{(1)}  ( E ) ]^{-1} \cdot \Sigma_1^{-1} [ \Phi] \cdot 
 \Omega_2 [ \Phi ]  \cdot [ \Upsilon^{(2)}  ( E ) ]  
 \;\;\;\; , 
 \label{ainv.c1}
 \eneq
 \noindent
for the p-wave hybrid ring, and

 \beq
 {\bf M}_s  =  [ \tilde{\Upsilon}^{(2)}  ( E ) ]^{-1} 
\cdot  \tilde{\Sigma}_2^{-1}  [ \Phi] \cdot \tilde{\Omega}_1 [ \Phi ]  \cdot [ 
\tilde{\Upsilon}^{(1)}  ( E ) ] 
\cdot {\bf T}_1 ( E  ; \ell_1)  
  \cdot [ \tilde{\Upsilon}^{(1)}  ( E ) ]^{-1} \cdot \tilde{\Sigma}_1^{-1} [ \Phi] \cdot 
 \tilde{\Omega}_2 [ \Phi ]  \cdot [ \tilde{\Upsilon}^{(2)}  ( E ) ] 
 \:\:\:\: , 
 \label{ainv.c2}
 \eneq
 \noindent
for the s-wave hybrid ring. From Eqs.(\ref{ainv.c1},\ref{ainv.c2}), we see that 
the current in the p-wave and the s-wave hybrid ring, $I_{ p , s } [ \Phi ]$, 
can respectively be written as 
 
 \beq
I_{ p ,  s }  [ \Phi ] = -  \frac{1}{2 \pi i } \: 
\int_\Gamma \: d E \: \partial_\Phi \ln [ \Psi_{ p , s }  [ E ; \Phi ] ] 
=   -  \frac{1}{4 \pi i } \:\int_\Gamma  \: d E \: \partial_\Phi \Biggl\{  \ln {\rm det} [ {\bf T}_2^{-1} ( \epsilon ; \ell_2 ) 
- {\bf M}_{ p , s }  ] -  [ \ln {\rm det} [ {\bf T}_2^{-1} ( \epsilon ; \ell_2 ) 
- {\bf M}_{ p , s }  ] ]^* \Biggr\}
\;\;\;\; , 
\label{ainv.c3}
\eneq
\noindent
where we have used the reality of the persistent 
current  to go through the last step in Eq.(\ref{ainv.c3}). In order 
to systematically take the $\ell_2 \to \infty$-limit, we recall that 
one has eventually to deform the integrals over $\Gamma$ in Eq.(\ref{ainv.c3}) 
into integrals over the imaginary axis, which corresponds to $ E \to 
i \omega$. Along the imaginary axis, from the dispersion relations for 
particle- and hole-like excitations within the superconducting region, 
one obtains that the corresponding momenta are defined by 
 
 \begin{eqnarray}
  \cos [ k_p^{(2)} ] &=& - \frac{\mu_2 w_2}{2 ( w_2^2 - \Delta_2^2 ) } - \frac{i}{2} \sqrt{\frac{\omega^2 + \Delta_w^2}{w_2^2 - \Delta_2^2}} 
  \nonumber \\
  \cos [ k_h^{(2)} ] &=& - \frac{\mu_2 w_2}{2 ( w_2^2 - \Delta_2^2 ) } + \frac{i}{2} \sqrt{\frac{\omega^2 + \Delta_w^2}{w_2^2 - \Delta_2^2}} 
  \;\;\;\; , 
  \label{inv_dr}
 \end{eqnarray}
\noindent
in the p-wave case, and 

 \begin{eqnarray}
  \cos [ k_p^{(2)} ] &=& - \frac{\mu_2 }{2   w_2 } - \frac{i}{2} \sqrt{\frac{\omega^2 + \Delta_2^2}{w_2^2 }} 
  \nonumber \\
  \cos [ k_h^{(2)} ] &=& - \frac{\mu_2 }{2   w_2 } + \frac{i}{2} \sqrt{\frac{\omega^2 + \Delta_2^2}{w_2^2}} 
  \;\;\;\; , 
  \label{inv_dr2}
 \end{eqnarray}
\noindent
 in the s-wave case. To solve Eqs.(\ref{inv_dr}) we therefore set 
 
 \beq
 k_p^{(2)} = \frac{\pi}{2} +  q_p \;\;\; , \;\; k_h^{(2)} = \frac{\pi}{2} + q_p^*
 \:\:\:\: , 
 \label{xx.1}
 \eneq
 \noindent
 with 
 
 \beq
 \sin [ q_p ] =  \frac{\mu_2 w_2}{2 ( w_2^2 - \Delta_2^2 ) } 
 + \frac{i}{2} \sqrt{\frac{\omega^2 + \Delta_w^2}{w_2^2 - \Delta_2^2}} 
 \;\;\;\; , 
 \label{xx.2}
 \eneq
 \noindent
 while, to solve Eqs.(\ref{inv_dr2}), we   set

 \beq
 k_p^{(2)} = \frac{\pi}{2} +  q_s \;\;\; , \;\; k_h^{(2)} = \frac{\pi}{2} + q_s^*
 \:\:\:\: , 
 \label{xx.3}
 \eneq
 \noindent
 with 
 
 \beq
 \sin [ q_s ] = \frac{\mu_2 }{2   w_2 } + \frac{i}{2} \sqrt{\frac{\omega^2 + \Delta_2^2}{w_2^2}} 
 \;\;\;\; .
 \label{xx.4}
 \eneq
 \noindent
From the explicit formula for ${\bf T}_{2 ; ( p , s )}^{-1} ( E \to i \omega ; \ell_2 )$ along 
the imaginary axis in the p-wave and in the s-wave case, respectively given by

\beq
{\bf T}_{2 ; ( p , s )}^{-1} ( E \to i \omega ; \ell_2 ) = 
\left[ \begin{array}{cccc}
                                                     i^{- \ell_2} e^{- i  q_{p , s}^{(2)} \ell_2 } 
                                                     & 0 & 0 & 0 \\ 
                                                     0 &    i^{ \ell_2} 
                                                     e^{ i  q_{p , s}^{(2)} \ell_2 } & 0 & 0 \\
       0 & 0  &    i^{ \ell_2} e^{  i [ q_{p , s}^{(2)}   ]^*   \ell_2 } & 0   \\   
       0 & 0 & 0 &    i^{ - \ell_2} e^{ -  i [ q_{p , s}^{(2)} ]^*  \ell_2 }                                     
                                                    \end{array} \right]
                                                    \:\:\:\: ,
                                                    \label{xx.7}
                                                    \eneq
\noindent
we may readily compute  the integrals in Eq.(\ref{ainv.c3}) in the limit $\ell_2 \to \infty$, obtaining   

 \beq
I_{ p , s }  [ \Phi ] =    \frac{1}{2 \pi  } \: \int_{ - \infty}^\infty  \: 
d \omega  \: \partial_\Phi \ln {\cal G}_{ p , s }  ( E \to i \omega ) 
\;\;\;\; . 
\label{xx.8a}
\eneq
\noindent
with  ${\cal G}_{ p ,s }  ( E )  = {\bf M}_{( p , s ); (2,2) } 
(E ) {\bf M}_{( p , s ); (4 , 4)} (E )   - 
{\bf M}_{( p , s ); (2,4) }  (E ) {\bf M}_{( p , s ); (4,2) }  (E )$ 
and assuming (as done in Ref.[\onlinecite{giu_af1,giu_af2}]) that 

\begin{itemize}
\item All the poles of ${\cal G} ( E )$ lie over the real axis;

 \item ${\cal G} ( E )$ is real if $E $ lies over the real axis (and 
 does not coincide with a pole of ${\cal G}$),
\end{itemize}
\noindent 
Eq.(\ref{xx.8a}) yields the dc-Josephson current in the infinite-$\ell_2$ limit, 
in which the ring can be regarded  as an idealized model for an SNS-junction. 
In the specific case of s-wave superconductors, Eq.(\ref{xx.2}) has been derived
in Ref.[\onlinecite{giu_af1}] for a single-channel junction starting from the $S$-matrix approach to 
effectively one-dimensional SNS-junctions \cite{been_0}, and generalized in Ref.[\onlinecite{giu_af2}] 
to a multi-channel junction. In fact, a comparison between
Eq.(\ref{xx.8a}) and Eqs.(7,9) of Ref.[\onlinecite{giu_af1}] also 
clarifies the identification between ${\bf M}_{p, s }$ in 
Eqs.(\ref{ainv.c1},\ref{ainv.c2}) and the transfer matrix for the whole
SNS-junction, as introduced and discussed in 
Ref.[\onlinecite{giu_af1}] for the s-wave case. After resorting to the 
effective SNS-junction model, at a second stage one may implement the 
technique developed in Refs.[\onlinecite{giu_af1,giu_af2}] to write 
$ I [ \Phi ]$ in a systematic expansion in inverse powers of $\ell_1$. 
Basically, one considers that, because one has

\beq
{\bf T}_1  ( E \to i \omega ; \ell_2 ) = \left[ \begin{array}{cccc}
                                                     i^{ \ell_1} e^{ i  q_n^{(1)} \ell_1 } & 0 & 0 & 0 \\ 
                                                     0 &    i^{ - \ell_1} e^{-  i  q_n^{(1)} \ell_1 } & 0 & 0 \\
       0 & 0  &  i^{ - \ell_1} e^{   - i [ q_n^{(1)}   ]^*   \ell_1 } & 0   \\    0 & 0 & 0 &    i^{  \ell_1} e^{   i [ q_n^{(1)} ]^*  \ell_1 }                                     
                                                    \end{array} \right]
                                                    \:\:\:\: , 
                                                    \label{xx.7bis}
                                                    \eneq
\noindent
with 
 \beq
 \sin [ q_n^{(1)}  ] = \frac{\mu_1 }{2   w_1 } + \frac{i}{2}  \frac{\omega  }{w_1} 
 \;\;\;\; ,
 \label{xx.4bis}
 \eneq
 \noindent
then, only low-$ | \omega |$ regions do effectively contribute to the integral in Eq.(\ref{xx.8a}).
As a result, one may first of all approximate $q_n^{(1)}  \approx \bar{q} + i \sigma ( \omega )$, with 

\begin{eqnarray}
 \sin ( \bar{q}  ) &=&  \frac{\mu_1 }{2   w_1 }  \nonumber \\
 \sigma ( \omega )  &=&    \frac{\omega  }{2 w_1 \cos (\bar{q})} 
 \;\;\;\; , 
 \label{xx.9}
\end{eqnarray}
\noindent
then, in integrating Eq.(\ref{xx.8a}), one may rescale $\omega \to \omega  \ell_1$ and eventually
set the rescaled $\omega$ at 0 in all the contributions to the argument of the integral in 
which $\omega$ appears divided by $\ell_1$ as $\omega / \ell_1$.  Going along this procedure, 
one may compute the leading contribution to the current $({\cal O} ( \ell_1^{-1} ))$ by trading 
the original model Hamiltonian for a reduced boundary model, such as the one presented in 
Ref.[\onlinecite{acz}] for the s-wave superconductors and the one used in Ref.[\onlinecite{giuaf_3}]
for the p-wave superconductor, which allows for recovering simple, closed-form analytical 
formulas for $ I [ \Phi ]$. 

 \subsection{The limit of long normal region}
 \label{long_normal}

We now discuss the complementary limit of a long normal region, with $\ell_S$ less than, or comparable to, 
the coherence length of the superconducting region $\xi_0$. We first discuss the general formula and then
consider the case of a hybrid s-wave superconducting ring as a specific example. In order to address
the large-$\ell_N$-limit, let us first of all rewrite 
${\rm det} \{  {\cal M}_{ p - wave \: (s - wave ) } [ E ; \Phi ; \ell_1 ; \ell_2 ] - {\bf I} \}$ 
as 

\beq
 {\rm det} \{  {\cal M}_{ p - wave  \; (s - wave) } [ E ; \Phi ; \ell_1 ; \ell_2 ] - {\bf I} \}  = c_{p , s }   \: {\rm det} \{  {\bf T}_1 ( E ; \ell_1 ) - 
 {\bf K}_{ p ,s } [ E ; \Phi ; \ell_2 ] \}
 \:\:\:\: ,
 \label{lnr}
\eneq
\noindent
with $c_{ p , s }$ being constants independent of $\Phi$, and 

\begin{eqnarray} 
&& {\bf K}_{ p   } [ E ; \Phi ; \ell_2 ]  =  [ \Upsilon^{(1)}   ( E ) ]^{-1} \cdot \Omega_1^{-1}  [ \Phi ]  \cdot 
\Sigma_2   [ \Phi] \cdot  [  \Upsilon^{(2)}  ( E ) ]  \cdot {\bf T}_2^{-1} ( E ; \ell_2 ) \cdot   
 [ \Upsilon^{(2)}  ( E ) ]^{-1}  \cdot \Omega_2^{-1}  [ \Phi ]  \cdot 
\Sigma_1   [ \Phi] \cdot  [ \Upsilon^{(1)}  ( E ) ]   \nonumber \\
&& {\bf K}_{ s  } [ E ; \Phi ; \ell_2 ]  =  [ \tilde{\Upsilon}^{(1)}   ( E ) ]^{-1} \cdot \tilde{\Omega}_1^{-1}  [ \Phi ]  \cdot 
\tilde{\Sigma}_2   [ \Phi] \cdot  [ \tilde{\Upsilon}^{(2)}  ( E ) ]  \cdot {\bf T}_2^{-1} ( E ; \ell_2 ) \cdot   
 [ \tilde{\Upsilon}^{(2)}  ( E ) ]^{-1}  \cdot \tilde{\Omega}_2^{-1}  [ \Phi ]  \cdot 
\tilde{\Sigma}_1   [ \Phi] \cdot  [ \tilde{\Upsilon}^{(1)}  ( E ) ]   
 \:\:\:\: . 
 \label{lnr.2}
 \end{eqnarray}
 \noindent
Equations (\ref{lnr.2}) correspond to the standard identification we have employed so far, that is, 
region-1 has to be  identified with the normal region and region-2 with  the (either p-wave, or s-wave) 
superconducting region. Therefore, $\ell_1 = \ell_N$. Now, in order to recover the large-$\ell_N$-limit, 
we strictly follow the derivation of Refs.[\onlinecite{giu_af1,giu_af2}], that is, once we have deformed the 
integration path to the imaginary axis, we assume that only low-$\omega ( = - i E )$ regions do effectively 
contribute the integral in Eq.(\ref{ra.11}). This allows us first of all to approximate the inverse
dispersion relation within the normal region as 

\beq
k_{ p , h } \approx k_F \pm i \lambda ( \omega ) 
\;\;\;\; , 
\label{lnr.3}
\eneq
\noindent
with $- 2 w_1 \cos ( k_F ) = \mu_1$ and $\lambda ( \omega ) = \frac{\omega}{2 w \sin (k_F)}$. On substituting 
Eqs.(\ref{lnr.3}) into Eqs.(\ref{lnr.2}), we may eventually rewrite Eq.(\ref{ra.11}) as 

\beq
I_{ p , s} 
[ \Phi ] = - \frac{2 e w_1 \sin (k_F)}{2 \pi \: \ell_1} \: \partial_\Phi \; \int_0^\infty \: \frac{d z}{z } \: \ln \{ \Xi_{ p , s }  [ z ; \Phi ;  \ell_2 ] \} 
\;\;\;\; , 
\label{lnr.4}
\eneq
\noindent
with 

\beq
 \Xi_{ p , s }  [ z ; \Phi ;  \ell_2 ] = {\rm det} \left\{ \left[ \begin{array}{cccc}
                                                         z & 0 & 0 & 0 \\ 0 & z^{-1} &0 & 0 \\ 0 & 0 & z & 0 \\ 0 & 0 & 0 & z^{-1} 
                                                        \end{array} \right] - 
  \left[ \begin{array}{cccc}
                                                         e^{ - i k_F \ell_1 }  & 0 & 0 & 0 \\ 0 & 
                                                         e^{ i k_F \ell_1 } &0 & 0 \\ 0 & 0 & e^{ i k_F \ell_1} & 0 \\ 0 & 0 & 0 & e^{ - i k_F \ell_1 }                                                                                                               
                                                        \end{array} \right] \cdot  {\bf K}_{ p  , s   } [ E = 0 ; \Phi ; \ell_2 ]                                                      
\right\}
\:\:\:\: . 
\label{lnr.5}
\eneq
\noindent
To  illustrate the effectiveness of our simplified Eqs.(\ref{lnr.4},\ref{lnr.5}) we now discuss the application to the 
case of a hybrid s-wave superconducting ring. For simplicity, we make the assumptions 
$w_1 = w_2 \equiv w$ and $\mu_1 = 0$. As a result, using Eqs.(\ref{lnr.4},\ref{lnr.5}), we obtain 
the simplified expression for the current

\beq
I  [ \Phi ] = - \frac{2 e w_1 \sin (k_F)}{2 \pi \: \ell_1} \: \partial_\Phi \; \int_0^\infty \: \frac{d z}{z } \: 
\left\{ \frac{ \partial_\Phi a [ \Phi ; \ell_1 ] ( z^3 + z ) + \partial_\Phi  b [ \Phi ; \ell_2 ] z^2}{
z^4 + 1 + a [ \Phi ; \ell_2 ] ( z^3 + z ) + b [ \Phi ; \ell_2 ] z^2 } \right\} 
\;\;\;\; , 
\label{lnr.6}
\eneq
\noindent
with $ a [ \Phi ; \ell_2 ] , b [ \Phi ; \ell_2 ]$ being long, though 
straightforward to derive, functions of the matrix elements of 
${\bf K}_{ s  } [ E = 0  ; \Phi ; \ell_2 ] $. Eventually, by means of 
simple manipulations Eq.(\ref{lnr.6}) can be expressed as a closed-form formula 
only of the four roots $z_j [ \Phi ; \ell_2 ]$ ($ j = 1 , \ldots , 4$) of the polynomial equation 
$z^4  + 1 + a [ \Phi ; \ell_2 ] ( z^3 + z  ) + b [ \Phi ; \ell_2 ] z^2 = 0$, which 
take the generic form 

\begin{eqnarray}
&& z_j [ \Phi ; \ell_2 ] = - \frac{a [ \Phi ; \ell_2] }{4} +u_j  \frac{\sqrt{8 + a^2 [ \Phi ; \ell_2 ] - 
4 b [ \Phi ; \ell_2]}}{4} \nonumber \\
& + & v_j  \frac{1}{2}\sqrt{- 2 + \frac{a^2 [ \Phi ; \ell_2]}{4}- b [ \Phi ; \ell_2 ] + 
u_j \frac{a^3 [ \Phi ; \ell_2] + 8 a [ \Phi ; \ell_2] - 4 a [ \Phi ; \ell_2 ] b [ \Phi ; \ell_2] }{2 \sqrt{8 -
a^2 [ \Phi ; \ell_2] - 4 b [ \Phi ; \ell_2 ] b [ \Phi ; \ell_2 ] }}
}
\;\;\;\; ,
\label{lnr.7}
\end{eqnarray}
 with $u_j , v_j = \pm 1$. Taking into account that $\prod_{ j = 1}^4 z_j [ \Phi ; \ell_2 ] = 1$, 
 one eventually obtains from Eq.(\ref{lnr.6})
 
 \beq
 I [ \Phi ] = - \frac{2 e w_1 \sin (k_F)}{4 \pi \ell_1} \partial_\Phi \{ \sum_{ j  = 1}^4 \ln^2 [ z_j [ \Phi ; \ell_2 ] ]\}
 \;\;\;\; . 
 \label{lnr.8}
 \eneq
 \noindent
 Just as in the case of a long SNS-junction \cite{giu_af1,giu_af2}, Eq.(\ref{lnr.8}) only involves
 data at the Fermi level. This is an additional example of the remarkable simplifications to 
 which our approach leads, in the large ring limit.

\section{Finite-temperature results}
\label{finite_T}

\begin{figure}
\includegraphics*[width=1\linewidth]{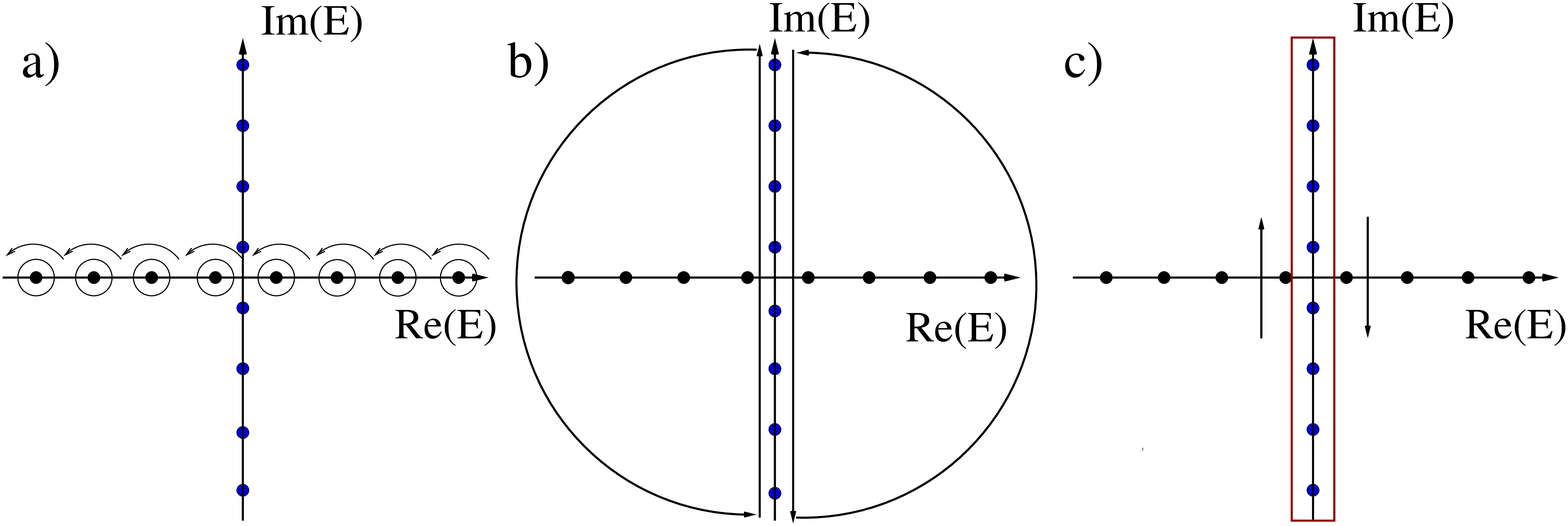}
\caption{Sequence of deformations in the integration path $\tilde{\Gamma}$ eventually 
allowing to express $ I [ \Phi ; T ]$ as a sum over the fermionic Matsubara frequencies
$\omega_m = 2 \pi T \left( m + \frac{1}{2} \right)$: \\
{\bf a)} The path $\tilde{\Gamma}$ obtained as the union of small circles, each one 
surrounding one, and only one,  energy
eigenvalue; \\
{\bf b)} The integral over $\tilde{\Gamma}$ can be deformed to an integral over the union of the
two closed path run through counterclockwise;  \\
{\bf c)} The integral over the two closed path in {\bf b)} is equal  to
the integral over a closed path surrounding the poles of the Fermi function ($i \omega_m$, displayed
as blue full circles in the figure), run through clockwisely. The corresponding 
over-all - sign is eventually canceled by the -1 at the residue of the Fermi function at
$i \omega_m$.} \label{defor_2}
\end{figure}

It is not difficult to generalize our derivation to a system at temperature $T$ finite, 
though much lower than the critical temperature for the superconducting part of the 
ring. In fact, at finite $T$, Eq.(\ref{ra.11}) generalizes to Eq.(\ref{one}), with  
 
\beq
F [ \Phi ; T ] = \sum_E E f ( E ) 
\;\;\;\; , 
\label{ft.2}
\eneq
\noindent
and $ f ( E ) = [ 1 + e^{ E / T } ]^{-1}$ being the Fermi distribution 
function (having set the Boltzmann constant $k = 1$. Now, the integration path 
$\Gamma$ in Eq.(\ref{ra.9}) must be replaced with the integration path 
$\tilde{\Gamma}$ obtained as the union of small circles $\Gamma_n$, each 
one surrounding once one, and only one, energy eigenvalue. As illustrated in 
Fig.\ref{defor_2}, $\tilde{\Gamma}$ can be deformed to a path obtained as the union
of small circles, each one surrounding once one, and only one, pole of the Fermi function,
that is, a fermionic Matsubara frequency times the imaginary unit $i$, $i \omega_m = 2 \pi T i \left( m + \frac{1}{2} \right)$.
As a result, the finite-$T$ current can be presented as 

\beq
I [ \Phi ; T ] = - e T \sum_{  \omega_m } \partial_\Phi \{ {\rm det} [  {\cal M} [ i \omega_m ; \Phi ] -  {\bf I} ] \}
\;\;\;\; , 
\label{ft.3}
\eneq
\noindent
that is the appropriate generalization of Eq.(\ref{ra.11}) to the finite-$T$ case.

\noindent

\section{Discussion and Conclusions}
\label{conclusions}

In this paper we have presented a technique to exactly compute the zero-temperature persistent current    across 
an HR pierced by a magnetic flux $\Phi$ as a single integral 
of a known function of the system's parameters. Our approach makes use of the 
properties of the transfer matrix of the ring, which allows us to circumvent technical 
difficulties associated with the secular equation for the energy eigenvalues of the system.
A straightforward generalization of the zero-temperature formalism allows us to also compute the current 
in a ring at a temperature $T$ finite, though much lower than the superconducting gap. 
While in general one may readily numerically compute the integral/sum yielding the current 
at a given value of the flux $\Phi$, a remarkable simplification takes place in the limit
of a large ring size, where resorting to a systematic expansion in inverse powers of the 
ring length allows for deriving the current in analytic closed-form formulas, applicable to 
a number of cases of physical interest. 

As an example of application of our technique, we exactly compute the persistent current through
a p-wave superconducting-normal ring as well as in an s-wave superconducting-normal ring. As a side result, 
we recover at once  the crossover in the current periodicity, the 
effects of localized MFs in the p-wave case (including the signature of the topological 
phase transition), the large-size limit, etc., which were previously approximatively derived by means
of various approximation schemes, different from case to case \cite{pientka,buttiker,giu_af1,giu_af2}.

Throughout all the paper, we grounded our approach on a description of the ring in 
terms of a noninteracting lattice model Hamiltonian  for the normal part of the ring, 
in terms of a non-self consistent mean field Hamiltonian for the superconducting part. 
The system is assumed to be ''clean'', that is, no disorder effects are taken into account and 
no interaction between the electrons is assumed. As a further development of our research, we
plan to generalize our approach, so to introduce the effect of disorder by means of a
pertinently chosen impurity potential and by eventually ensemble averaging over 
the disorder realization \cite{RefD,RefE}, and the effect of the interaction, by adopting a pertinently 
adapted version of the fermion renormalization group approach introduced in Refs.[\onlinecite{frg_1,frg_2,frg_3}] for 
weakly interacting systems and generalized in Ref.[\onlinecite{giul_nava}] to the strongly interacting case. This might potentially provide an useful method to also spell out the dynamics of fractionalized excitations in correlated systems \cite{RefA,RefB,RefC}. Moreover, by simply sending to infinity the length of the superconducting part
of the ring, our approach provides an exact formula for the Josephson current 
across an SNS junction, which can be readily applied to cases hard to deal with 
using alternative techniques, such as in the case of the anomalous Josephson effect
in nanowires \cite{naza_1,naza_2,clgt}.

These, and other potentially interesting generalizations of our approach, do actually lie 
beyond the range of this paper, and we plan to address them as a future development of 
our work.

 \vspace{0.5cm}

We thank S. K. Maiti for a very useful correspondence exchange during the preparation
of this work. We acknowledge insightful discussions with L. Lepori and A. Tagliacozzo. 
 
\appendix

\section{Exact wavefunctions for sub-gap states at a finite-length 
Kitaev chain}
\label{exw}

In this section, we discuss in detail the derivation of the exact wavefunction 
of sub-gap states in an open Kitaev chain, of finite length $\ell_S$. To do so, we employ a 
simplified and pertinently adapted version of the solution of the Kitaev model 
with open boundary conditions \cite{giu_stt,fagotti}. While, for
finite-$\ell_S$, we find a single Dirac fermion level, which can be either empty, or
filled, with a corresponding finite energy gap $E_M$, as $\ell_S \to \infty$, the 
states become degenerate in energy at the Fermi level, and appropriate linear combinations
of the corresponding wavefunctions become localized at the two endpoints of the chain, 
eventually corresponding to the two zero-energy Majorana solutions of Kitaev's model
\cite{kita}. The starting point is the dispersion relation in Eqs.(\ref{pw.11})
which, for sub-gap solutions ($ | E | < \Delta_w$) yields the allowed values of 
the (complex conjugate) particle- and hole-momenta defined by 

\begin{eqnarray}
 \cos ( k_p ) &=& - \frac{w \mu}{2 ( w^2 - \Delta^2 ) } - \frac{i}{2}
 \sqrt{\frac{\Delta_w^2 - E^2}{w^2 - \Delta^2 }} \nonumber \\
  \cos ( k_h ) &=& - \frac{w \mu}{2 ( w^2 - \Delta^2 ) } + \frac{i}{2}
 \sqrt{\frac{\Delta_w^2 - E^2}{w^2 - \Delta^2 }}
 \;\;\;\; .
 \label{appe.1.1}
\end{eqnarray}
\noindent
Eqs.(\ref{appe.1.1}) are readily solved by setting $k_p = \pi - q_R + i q_I$ and 
$k_h = ( k_p )^* = \pi - q_R - i q_I$, with 

\begin{eqnarray}
\cos ( q_R ) \cosh ( q_I ) &=&   \frac{w \mu }{2 ( w^2 - \Delta^2  ) }
\nonumber \\
\sin ( q_R ) \sinh ( q_I ) &=&  \frac{1}{2}
 \sqrt{\frac{\Delta_w^2 - E^2}{w^2 - \Delta^2 }} 
 \:\:\:\: . 
 \label{appe.1.2}
\end{eqnarray}
\noindent
As a result, the general formula for a subgap solution will be given by 

\begin{eqnarray}
\left[ \begin{array}{c}
        u_j \\ v_j 
       \end{array} \right] &=& (-1)^j \Biggl\{ A_{ (p , + ) } 
       \left[ \begin{array}{c}
               u \\ v 
              \end{array} \right] e^{ - i q_R j } e^{ - q_I j } 
              + A_{ (p , - ) } 
       \left[ \begin{array}{c}
               u \\ - v 
              \end{array} \right] e^{ i q_R j } e^{   q_I j } \nonumber \\
&+& A_{ (h , + ) } 
       \left[ \begin{array}{c}
               u^* \\  v^* 
              \end{array} \right] e^{  i q_R j } e^{ - q_I j } 
              + A_{ (h , - ) } 
       \left[ \begin{array}{c}
               u^* \\ - v^* 
              \end{array} \right] e^{ - i q_R j } e^{   q_I j } \Biggr\}
       \:\:\:\: , 
       \label{appe.1.3}
\end{eqnarray}
\noindent
with $u , v$ solutions of the algebraic system 

\begin{eqnarray}
&& \{ E + 2 w \cos (  k_p ) + \mu \} u - 2 i \Delta \sin ( k_p ) v = 0 \nonumber \\
&& 2 i \Delta \sin ( k_p ) u +  \{ E - 2 w \cos (  k_p ) - \mu \} v = 0
\:\:\:\: . 
\label{appe.1.4}
\end{eqnarray}
\noindent
The actual energy eigenstates are determined as nontrivial solutions such 
as the one in Eq.(\ref{appe.1.3}) satisfying the boundary conditions 
on the ''ghost sites'' given by $ \left[ \begin{array}{c}
          u_0 \\ v_0 
         \end{array} \right] =   \left[ \begin{array}{c}
          u_{\ell + 1}  \\ v_{\ell + 1} 
         \end{array} \right]  = 0$. It is therefore straightforward to
verify that this implies the secular equation for the energy eigenvalues 
given by 

\beq
[\Im m ( u v^* ) ]^2 \sinh^2 [ q_I ( \ell + 1 ) ] = [ \Re e  ( u v^* )]^2 
\sin^2 [ q_R ( \ell + 1 ) ] 
\:\:\:\: . 
\label{appe.1.5}
\eneq
\noindent
Eqs.(\ref{appe.1.3},\ref{appe.1.4},\ref{appe.1.5}) can be used to 
estimate the energy of the sub-gap levels and the corresponding 
wavefunction for any value of $\mu$. Indeed, we use them to
numerically estimate the energy gap and to accordingly infer the 
overlap scale between the localized MFs at given 
values of the system's parameters. Specifically, of one is only
interested in the low-energy physics of a finite-size one-dimensional
topological superconductor coupled to normal conductors at each of 
its endpoints, the whole topological superconductor can be 
traded for an effective Hamiltonian involving only the low-energy
sup-gap degrees of freedom discussed above, with parameters effectively
determined by the actual system parameters. To illustrate how the 
procedure works, let us focus onto the simple case   $\mu = 0$, $\ell$ even. In
this case, $k_p$ and $k_h$ are simply given by 
 
\begin{eqnarray}
k_p &=&  \frac{\pi}{2} + i  \lambda ( E )  \nonumber \\
k_h &=&  \frac{\pi}{2} -  i  \lambda ( E ) 
\:\:\:\: , 
\label{mfo.3}
\end{eqnarray}
\noindent
with 

\beq
\sinh [ \lambda ( E ) ] = \frac{1}{2} \: 
\sqrt{\frac{ 4 \Delta^2 - E^2}{w^2 - \Delta^2 } } 
\:\:\:\: . 
\label{mfo.4}
\eneq
\noindent
As a result, Eq.(\ref{appe.1.3}) can now be presented as 

\begin{eqnarray}
\left[ \begin{array}{c}
        u_j \\ v_j 
       \end{array} \right] &=& 
A_{ ( p , + ) } \left[ \begin{array}{c}
                    u  \\ v 
                   \end{array} \right] i^{ j} e^{ - \lambda (E) j } + 
 A_{ ( p , - ) } \left[ \begin{array}{c}
                    u \\ - v 
                   \end{array} \right] i^{-j} e^{  \lambda (E) j } \nonumber \\
                   &+& 
 A_{ (  h , + ) }  \left[ \begin{array}{c}
               u^* \\  v^* 
              \end{array} \right] i^{-j} e^{ - \lambda (E) j } +                    
  A_{ ( h , - ) } \left[ \begin{array}{c}
               u^* \\ - v^* 
              \end{array} \right] i^{ j} e^{  \lambda (E) j }       
\:\:\:\: , 
\label{mfo.14}
\end{eqnarray}
\noindent
with 
\beq
\left[ \begin{array}{c}
        u_p   \\ v_p  
       \end{array} \right] = 
       \frac{1}{\sqrt{2}} \: \left[ \begin{array}{c} 
             e^{ \frac{i }{2} {\rm sgn} ( E ) \vartheta  }    \\ 
               i {\rm sgn} ( E )  e^{ -  \frac{i }{2} {\rm sgn} ( E ) \vartheta  } 
              \end{array} \right] \;\;\; ; \;\; 
\left[ \begin{array}{c}
        u_h  \\ v_h 
       \end{array} \right] = 
       \frac{1}{\sqrt{2}} \: \left[ \begin{array}{c} -  i{\rm sgn} ( E ) 
       e^{ -  \frac{i }{2}{\rm sgn} ( E ) \vartheta  } \\ 
             e^{ \frac{i }{2}{\rm sgn} ( E ) \vartheta  } 
              \end{array} \right] 
 \:\:\:\: ,
  \label{mfo.13} 
 \eneq
 \noindent
and 

\beq
 \vartheta = {\rm atan}  \left[ \frac{ 2 w \sinh [ \lambda (E ) ] }{ | E | } \right]
 \:\:\:\: . 
 \label{appe.1.6}
 \eneq
 \noindent
It is straightforward, though tedious, to show that the secular equation
for the sub-gap energy eigenvalues is now given by

\beq
 ( \ell + 1 ) 
\lambda (E) = \pm \sinh^{-1} \left[ \frac{2 w }{E} \sinh [ \lambda ( E ) ] \right]  
\:\:\:\: , 
\label{mfo.17}
\eneq
\noindent
which is   solved by setting 

\beq
| E | = \epsilon =  2 \Delta \left[ \frac{\cosh [ \lambda (E)] }{ \cosh [ ( \ell + 1 ) \lambda [E ]] } \right] 
\approx 2 \Delta e^{ - \ell \lambda (E) } \approx 2 \Delta \exp \left\{ - \ell \sinh^{ - 1} 
\left[ \frac{\Delta}{\sqrt{w^2 - \Delta^2} } \right] \right\}
\:\:\:\: . 
\label{mfo.18}
\eneq
\noindent
Therefore, from Eq.(\ref{mfo.18}) we readily estimate that the 
hybridization length scale between the MFs, $\ell_M$, 
is given by 

\beq
\ell_M \sim \left\{ \sinh^{ - 1} 
\left[ \frac{\Delta}{\sqrt{w^2 - \Delta^2} } \right] \right\}^{-1}
\:\:\:\: . 
\label{mfo.19}
\eneq
\noindent
From Eq.(\ref{appe.1.3}) we may therefore construct the wavefunctions corresponding to 
the positive- and to the negative-energy sub-gap solutions. As a result, 
one obtains for the positive- sub-gap energy solution of the BdG equations

\begin{eqnarray}
\left[ \begin{array}{c}
        u_j \\ v_j 
       \end{array} \right]_{ + } &=&  c \Biggl\{ 
        e^{ \frac{\xi}{2}}   \: \left[ \begin{array}{c} 
              e^{ \frac{i }{2}   \vartheta  }  \\ 
               i  e^{- \frac{i }{2}   \vartheta  } 
              \end{array} \right]  i^{ j} e^{ - j \lambda (E) } +i     
              e^{ - \frac{\xi}{2}}  \:
 \left[ \begin{array}{c} 
              e^{ \frac{i }{2}   \vartheta  }  \\ 
              - i  e^{- \frac{i }{2}   \vartheta  } 
              \end{array} \right] i^{-j} e^{  j \lambda (E) }             
\nonumber \\
&+&       e^{ \frac{\xi}{2}}  \:  \left[ \begin{array}{c} 
             -i  e^{ - \frac{i }{2}   \vartheta  }  \\ 
               - e^{ \frac{i }{2}   \vartheta  } 
              \end{array} \right]   i^{ - j } e^{ - j \lambda (E) } 
             -i    e^{ - \frac{\xi}{2}}  \: 
\left[ \begin{array}{c} 
                - i e^{ - \frac{i }{2}   \vartheta  }  \\ 
               e^{ \frac{i }{2}   \vartheta  } 
              \end{array} \right]   i^{    j } e^{  j \lambda (E) }  \Biggr\} 
\:\:\:\: , 
\label{mfo.22}
\end{eqnarray}
\noindent
with 

\beq
c = \frac{1}{4} \: \sqrt{\frac{ 2 \sinh [ \lambda (E)]}{ \sinh [ \xi (E) - \lambda (E) ]
}}
\:\:\:\: , 
\label{mfo.23X}
\eneq
\noindent
and $\xi (E) =  \sinh^{-1} \left[ \frac{2 w }{E} \sinh [ \lambda ( E ) ] \right]$.   
Similarly,
one obtains for the negative-energy sub-gap solution of the BdG equations

\begin{eqnarray}
\left[ \begin{array}{c}
        u_j \\ v_j 
       \end{array} \right]_{  - } &=&  - c \Biggl\{ 
          e^{ \frac{\xi}{2}}   \: \left[ \begin{array}{c} 
              e^{ - \frac{i }{2}   \vartheta  }  \\ 
              - i  e^{  \frac{i }{2}   \vartheta  } 
              \end{array} \right]  i^{ j} e^{ - j \lambda (E) } - i     
              e^{ - \frac{\xi}{2}}  \:
 \left[ \begin{array}{c} 
              e^{ -\frac{i }{2}   \vartheta  }  \\ 
                i  e^{ \frac{i }{2}   \vartheta  } 
              \end{array} \right] i^{-j} e^{  j \lambda (E) }             
\nonumber \\
&+&       e^{ \frac{\xi}{2}}  \:  \left[ \begin{array}{c} 
                i  e^{   \frac{i }{2}   \vartheta  }  \\ 
               -  e^{ - \frac{i }{2}   \vartheta  } 
              \end{array} \right]   i^{ - j } e^{ - j \lambda (E) } 
             + i    e^{  - \frac{\xi}{2}}  \: 
\left[ \begin{array}{c} 
                i e^{   \frac{i }{2}   \vartheta  }  \\ 
                 e^{ - \frac{i }{2}   \vartheta  } 
              \end{array} \right]   i^{  j } e^{  j \lambda (E) }  \Biggr\} 
\:\:\:\: .
\label{mfo.22x}
\end{eqnarray}
\noindent
From Eqs.(\ref{mfo.22},\ref{mfo.22x}), we therefore find that the 
eigenmodes corresponding to the $\pm$ solutions are respectively given by

\begin{eqnarray}
 \Gamma_{  + } &=& \sum_{ j = 1}^\ell \{ [ u_j ]^*_{  + } c_j + 
 [ v_j ]^*_{  + } c_j^\dagger \} 
 \nonumber \\
  \Gamma_{ - } &=& \sum_{ j = 1}^\ell \{ [ u_j ]^*_{   - } c_j + 
 [ v_j ]^*_{  - } c_j^\dagger \} 
 \:\:\:\: ,
 \label{mfo.24}
\end{eqnarray}
\noindent
which shows that, as expected, one recovers the relation

\beq
\Gamma_{+} = \Gamma_{  - }^\dagger \equiv \Gamma
\:\:\:\: . 
\label{mfo.25}
\eneq
\noindent
On inverting Eqs.(\ref{mfo.24}) and truncating the mode expansion of the real space 
lattice operators by retaining low-energy modes only,  we
therefore get 

\begin{eqnarray}
 c_j & \approx & [u_j]_{  + } \Gamma + [v_j ]^*_{ + } \Gamma^\dagger 
 \nonumber \\
 c_j^\dagger &\approx & [v_j]_{ + } \Gamma + 
 [u_j ]^*_{ + } \Gamma^\dagger 
 \:\:\:\: . 
 \label{mfo.26}
\end{eqnarray}
\noindent
Now, on calling $\{ d_j , d_j^\dagger \}$ and $\ell_N$ the length of the normal part of the ring, the 
fermion operators on the normal side, one may rewrite the tunneling contribution to the 
Hamiltonian in Eq.(\ref{pw.1}) as  
\beq
 H_\tau = -  \tau \{ [ c_1^\dagger d_{ \ell_N} + d_1^\dagger c_{ \ell_N } ] e^{ \frac{i}{4} \Phi } 
 + [ d_{ \ell_N}^\dagger c_1 + c_{ \ell }^\dagger d_1 ] e^{  - \frac{i}{4} \Phi }  \}
 \:\:\:\: . 
 \label{mfo.27}
\eneq
\noindent
Using the truncated expansions in Eqs.(\ref{mfo.26}) and the explicit form 
of the wavefunctions evaluated at   $j=1 , \ell$, one eventually approximates
Eq.(\ref{mfo.27}) as 

\beq
 H_\tau =  t_L \{ \gamma_L [  e^{ \frac{i}{4} \Phi} d_{\ell_N}  - e^{ - \frac{i}{4} \Phi}  d_{\ell_N}^\dagger ]  \} 
 + i t_R \{  \gamma_R [ e^{ - \frac{i}{4} \Phi}  d_1 + e^{ \frac{i}{4} \Phi}  d_1^\dagger ] \}  
 \:\:\:\: , 
 \label{mfo.31}
\eneq
\noindent
with  

\begin{eqnarray}
 \gamma_L &=& e^{ - i \frac{\pi}{4} }  \Gamma +  e^{ i \frac{\pi}{4} }  \Gamma^\dagger  \nonumber \\
 \gamma_R &=& -i \{ e^{ - i \frac{\pi}{4} }  \Gamma -  e^{ i \frac{\pi}{4} }  \Gamma^\dagger   
 \}
 \:\:\:\: .
 \label{mfo.32}
\end{eqnarray}
\noindent
and $t_L = t_R = \Upsilon \tau$, with

\begin{eqnarray}
 u_{1 , + } &=& - e^{ - i \frac{\pi}{4} } \Upsilon   \nonumber \\
  v_{ 1 , + } &=& - e^{ - i \frac{\pi}{4} } \Upsilon \nonumber \\
   u_{ \ell  , + } &=&  - e^{ - i \frac{\pi}{4} } \Upsilon  \nonumber \\
   v_{ \ell  , + } &=&  e^{ - i \frac{\pi}{4} } \Upsilon 
  \;\;\;\; . 
  \label{mfo.29}
\end{eqnarray}
\noindent
Finally, to recover the energy bias between the Dirac modes, we add
a term of the form 

\beq
H_\Gamma = 2  \epsilon \{ 2 \Gamma^\dagger \Gamma - 1 \} = 
- 2 \epsilon  i \gamma_L \gamma_R
\:\:\:\: . 
\label{mfo.32a}
\eneq
\noindent
In general, $t_L , t_R$ are smooth functions of $\mu$. The dependence of $\epsilon$ on 
$\mu$ can be inferred by numerically solving Eq.(\ref{appe.1.5}), as we did in 
the main text, to also derive the dependence of $\ell_M$ on the chemical potential.

\bibliography{bibhybrid_resub}

\end{document}